\documentclass[preprints,article,accept,pdftex,moreauthors]{Definitions/mdpi} 
\firstpage{1} 
\pubvolume{1}
\issuenum{1}
\articlenumber{0}
\pubyear{2022}
\copyrightyear{2022}
\datereceived{} 
\dateaccepted{} 
\datepublished{} 
\hreflink{https://doi.org/} % If needed use \linebreak
\pdfoutput=1

\usepackage{siunitx}
\usepackage{subcaption}
\Title{A Beam Monitor for Ion Beam Therapy based on HV-CMOS Pixel Detectors}

\Author{Matthias Balzer $^{1}$, Alexander Dierlamm $^{1,2,*}$\orcidA{}, Felix Ehrler $^{1,5}$\orcidB{}, Ulrich Husemann $^{2}$\orcidC{}, Roland Koppenhöfer $^{2}$\orcidG{}, Ivan Peri\'{c} $^{1}$, Martin Pittermann $^{2}$, Bogdan Topko $^{2}$\orcidE{}, Alena Weber $^{1,\ddag}$, Stephan Brons $^{3}$, Jürgen Debus $^{3,4,6}$\orcidF{}, Nicole Grau $^{5,6}$, Thomas Hansmann $^{3}$,  Oliver Jäkel $^{3,7}$, Sebastian Klüter $^{5}$\orcidD{}, Jakob Naumann $^{3}$}

\address{%
$^{1}$ \quad Institute for Data Processing and Electronics (IPE), Karlsruhe Institute of Technology, Germany\\
$^{2}$ \quad Institute of Experimental Particle Physics (ETP), Karlsruhe Institute of Technology, Germany\\
$^{3}$ \quad Heidelberg Ion-Beam Therapy Center (HIT), Department of Radiation Oncology, Heidelberg University Hospital, Germany\\
$^{4}$ \quad National Center for Tumor Diseases (NCT) and Heidelberg Institute of Radiation Oncology (HIRO), National Center for Radiation Research in Oncology (NCRO), Heidelberg University Hospital, Germany\\
$^{5}$ \quad Department of Radiation Oncology and Heidelberg Institute of Radiation Oncology (HIRO), National Center for Radiation Research in Oncology (NCRO) Heidelberg University Hospital, Germany\\
$^{6}$ \quad Clinical Cooperation Unit Radiation Oncology, German Cancer Research Center (DKFZ), Germany\\
$^{7}$ \quad Division of Medical Physics in Radiation Oncology, and Heidelberg Institute of Radiation Oncology (HIRO) and National Center for Radiation Research in Oncology (NCRO), German Cancer Research Center (DZFZ) Heidelberg, Germany\\

$^{\ddag}$ \quad now with Bosch AG
}

\corres{Correspondence: alexander.dierlamm@kit.edu}

\abstract{
Particle therapy is a well established clinical treatment of tumors. 
More than one hundred particle therapy centers are in operation world wide. 
The advantage of using hadrons like protons or carbon ions as particles for tumor irradiation is the distinct peak in the depth dependent energy deposition, which can be exploited to accurately deposit dose in the tumor cells. 
To guarantee this, high accuracy of monitoring and control of the particle beam is of utmost importance.
Before the particle beam enters the patient, it traverses a monitoring system which has to give fast feedback to the beam control system on position and dose rate of the beam while minimally interacting with the beam.
The multi-wire chambers mostly used as beam position monitor have their limitations when fast response time is required (drift time).
Future developments like MRI-guided ion beam therapy pose additional challenges for the beam monitoring system like tolerance of magnetic fields and acoustic noise (vibrations). 
Solid-state detectors promise to overcome these limitations and the higher resolution they offer can create additional benefits.
This article presents the evaluation of an HV-CMOS detector for beam monitoring, provides results from feasibility studies in a therapeutic beam and summarizes the concepts towards the final large-scale assembly and readout system.}


\begin{document}
%%%%%%%%%%%%%%%%%%%%%%%%%%%%%%%%%%%%%%%%%%
\section{Introduction}
In this section the concept of ion beam therapy is introduced followed by the requirements for a beam monitoring system and a description of a state-of-the-art system. 
This information is based on the beam application and monitoring system of the Heidelberg Ion beam Therapy center (HIT), but applies to similar ion beam therapy facilities, too. 
A short review of alternative beam monitoring concepts and devices follows. The section concludes with a description of the newly proposed system.

\subsection{Ion Beam Therapy}\label{sec:HIT}
\begin{figure}
    \centering
    \includegraphics[width=0.75\textwidth]{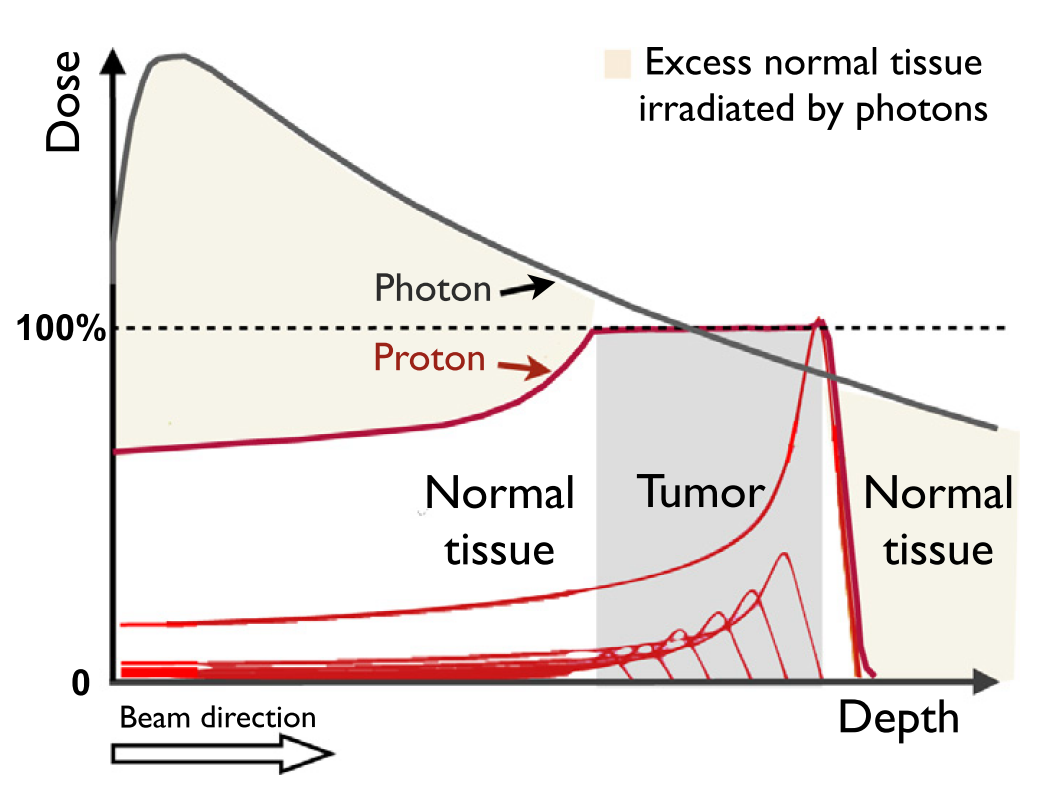}
    \caption{Energy deposition as a function of penetration depth for photons and protons. In contrast to photons, protons have a distinct Bragg peak. A spread-out proton peak can be achieved by superposition of energy depositions from proton beams of different energies to uniformly irradiate the entire tumor. From~\cite{royce_ProtonTherapyProstate_2019}.}
    \label{fig:bragg}
\end{figure}

\begin{figure}
    \centering
    \includegraphics[width=0.8\textwidth]{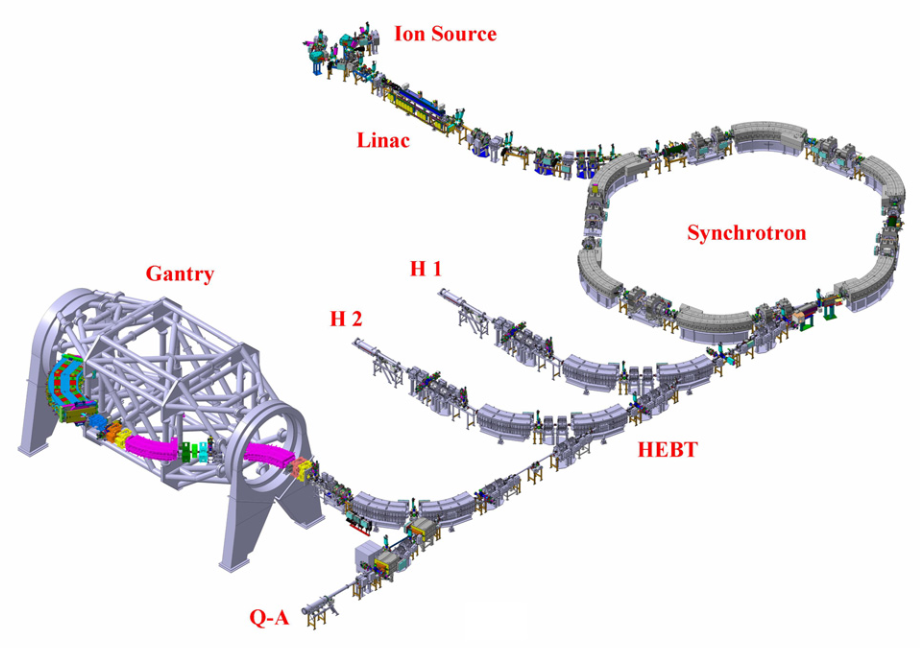}
    \caption{Overview of the accelerator complex (ion sources, linac, synchrotron and high energy beam transport line (HEBT)) and treatment rooms (H1 and H2 for the horizontal irradiation rooms, and Q-A for the quality assurance and experimental irradiation room) at HIT from~\cite{kleffner_HeidelbergIonTherapy_2009}. The large rotating gantry is on the left.}
    \label{fig:HIT_complex}
\end{figure}

\begin{figure}
    \centering
    \includegraphics[width=0.6\textwidth]{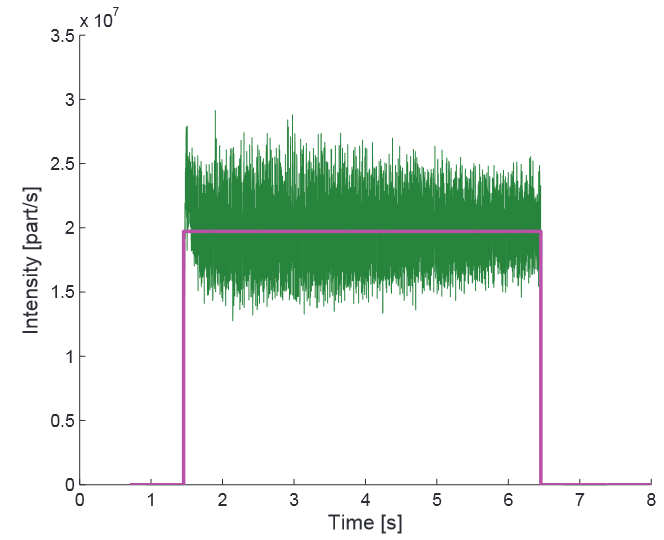}
    \caption{Time structure of a spill measured with an ionization chamber taken from~\cite{schomers_PatientspecificIntensitymodulationSlowly_2013}. 
    The magenta line represents the desired spill shape, while the green line is the actual measurement.
    This example is for a carbon ion beam with \SI{250}{MeV/u} and can be different for other beam settings.}
    \label{fig:HIT_spill}
\end{figure}

Today, ion beam therapy as a treatment against cancer is well established after a long history of developments~\cite{jakel_HistoryIonBeam_2022}. 
More than one hundred particle therapy facilities are being operated today\footnote{\url{https://www.ptcog.ch/index.php/facilities-in-operation}}. 
Protons or heavier ions can be well controlled both in energy by the accelerator and in direction by magnetic deflection in the beam delivery system.
The sharply peaking dose-versus-depth characteristic (Bragg peak) of these hadrons in tissue, as sketched in Figure~\ref{fig:bragg}, allows to control the depth of the deposited dose in the patient by the energy of the particles. 
Using a treatment plan with different raster points in different iso-energy slices (IES), highly conformal irradiation is possible, which maximizes the damage to the tumor while sparing healthy tissue.\\
HIT came into clinical operation in 2009~\cite{haberer_HeidelbergIonTherapy_2004}. 
It consists of a chain of ion accelerators designed for the scanning treatment technique which enables three-dimensional control of particle fluence and hence also dose.
The technique used here is called fluence controlled continuous spot scanning~\cite{haberer_MagneticScanningSystem_1993} and uses a continuously extracted beam, while scanning. 
The lateral motion is stopped at predefined spots until a predefined number of ions has been delivered, before the beam is quickly moving to the next spot.
Ions are extracted from one of the three ion sources, then accelerated in a linear accelerator (linac) up to \SI{7}{\mega\electronvolt\per u} and injected into the compact synchrotron with a circumference of about \SI{65}{\metre} after stripping off all electrons. 
The ions can reach energies of up to \SI{430}{\mega\electronvolt\per u}. 
The extracted energy is controlled directly by the synchrotron and may be varied from spill to spill to achieve the depth variation.
Clinically, carbon ions and protons are used.
\begin{table}[]
    \caption{Example parameter ranges describing therapeutic particle beams as they are implemented at HIT. Protons and carbon ions are selected here since they are most commonly used for treatments.}
    \centering
    \begin{tabular}{c|c|c}
        \toprule
         & Proton & Carbon \\ \midrule
        Energy (MeV/u) & \numrange{48}{221} & \numrange{88}{430}\\
        Intensity (\si{\per\second}) & \numrange{1.2e8}{3.2e9} & \numrange{5e6}{8e7}\\
        Width (FWHM in mm) & \numrange{8}{32} & \numrange{6}{12}\\ \bottomrule
    \end{tabular}
    \label{tab:beampara}
\end{table}
Table~\ref{tab:beampara} provides some basic beam parameters.\\
The ions are guided through the beam transport line to four irradiation rooms.
Two horizontal irradiation rooms and one room with a \SI{360}{\degree} rotating ion gantry are available for patient treatment~\cite{galonska_HitGantryCommissioning_2013}. 
A fourth room is used for quality assurance and research activities (QA room), and the measurements with a particle beam presented in Section~\ref{sec:Studies} have been performed there. 
All rooms are equipped for 3D scanning irradiations. 
An overview of the complex is shown in Figure~\ref{fig:HIT_complex}.
The accelerator control system can currently provide 255 different energy settings, 15 intensity steps and six spot sizes for each ion species; for practical reasons only a subset of the intensity steps and spot sizes is used.
The beam is produced in spills with lengths of up to \SI{5}{\s} and intensities up to \SI{3.2e9}{\per\second} for protons.
As the extracted beam intensity is not constant and may vary considerably during extraction, the monitor system is used to measure the number of delivered ions at any time. 
The beam intensity is controlled dynamically using a feedback loop which involves a monitor chamber~\cite{schomers_PatientspecificIntensitymodulationSlowly_2013}.
A measurement of the number of particles within a spill performed with an ionization chamber is shown in Figure~\ref{fig:HIT_spill}.
When a predefined number of ions is reached, the feedback loop to the scanner magnets triggers the movement of the beam to the next raster point~\cite{haberer_MagneticScanningSystem_1993}.
The spills can be aborted as soon as all raster points in one IES have been irradiated and before the energy has to be switched.
Fluence controlled continuous spot scanning requires a fast (typically within \SI{10}{\micro\second}) monitoring and control feedback loop, which is part of the Beam Application and Monitoring System described in Section~\ref{sec:BAMS}.\\
To fully benefit clinically from high precision dose delivery by scanned ion beams, image guidance in treatment position during the course of the treatment is mandatory. 
Real-time imaging can help locating the target volume in moving tissue and adapting the treatment plan during irradiation, but is currently not available for ion beams.
Unlike x-ray based computed tomography (CT), magnetic-resonance imaging (MRI) offers superior soft tissue contrasts and good discrimination between tumor and normal tissue. 
Moreover, it is not based on ionizing radiation and does not lead to additional imaging dose, which makes it suitable for continuous monitoring. 
For photon-based radiotherapy, the first hybrid treatment system came to the market in 2014 and combines \SI{0.35}{\tesla} MRI with intensity modulated radiotherapy.
While this initial system was based on Co-60 sources, linac based systems are available since 2016.
This system was also implemented for clinical use in Heidelberg~\cite{kluter_TechnicalDesignConcept_2019}. 
In contrast, MR-based image guidance for ion beam therapy is technologically even more challenging and its development is still in a very early phase. 
Recently, studies on the realization of MRI-guided proton therapy have started with several challenges ahead~\cite{hoffmann_MRguidedProtonTherapy_2020}. 
Within the ARTEMIS\footnote{Adaptive RadioThErapie Mit IonenStrahlen, funded by BMBF} project in Heidelberg, a demonstrator system is under development to implement MRI-guided ion beam delivery including online MR imaging of the target region during irradiation. 
Specifically, beam control and beam monitoring in the presence of an active MRI device need to cope with magnetic fields and acoustic noise, which specifically poses a problem for MWCs with thin windows (vibrations), and may thus call for alternative detector technologies.

\subsection{Beam Monitoring}\label{sec:BAMS}
The beam monitoring system needs to provide information about the position and particle number at each raster point and is placed close to the patient.
Additionally, control of the beam spot size is desirable, although this is not actively controlled during beam delivery.
The requirements for a next-generation beam monitoring system at HIT can be formulated as follows:
\begin{itemize}
    \item \textbf{Position measurement}: 
    The beam position can be calculated as the center of gravity of the induced signals. 
    The accuracy of the calculated position shall be better than \SI{200}{\micro\metre}.
    \item \textbf{Spot size measurement}: 
    The accuracy of the calculated beam spot width shall be better than 10\% for a Gaussian shape.
    \item \textbf{Fluence measurement}: 
    The number of particles needs to be counted for each raster point. The particle numbers (>\num{10000} carbon ions; >\num{200000} protons) shall be measured with 0.5\% accuracy.
    \item \textbf{Timing and latency}: 
    The position and spot size shall be measured within \SI{100}{\micro\second} and delivered to the fast control interface within \SI{100}{\micro\second} latency. An updated value shall be delivered to the fast interface every \SIrange{1}{2}{\micro\second} for the fluence measurement and every \SIrange{50}{100}{\micro\second} for the position and width measurement. 
    \item \textbf{Radiation tolerance}: 
    The system shall be operable for at least five years. 
    For the central position, where the maximum accumulated dose is expected (few mm$^2$), this corresponds to a particle fluence of \SI{3e14}{\per\square\centi\metre} with carbon ions and \SI{3e15}{\per\square\centi\metre} with protons, which falls off very fast with increasing radial distance. 
    This very localized radiation damage might be an additional challenge for the detector.
    The radiation damage deteriorates the performance of the system, which needs to be replace regularly.
    The regular replacement and recommissioning after five years operation shall not take longer than one day. 
    More frequent replacements shall not last longer than two hours.
    \item \textbf{Size of sensitive area}: 
    The overall active area shall be $\SI{25}{\centi\metre}\times\SI{25}{\centi\metre}$ to cover the treatment area of $\SI{20}{\centi\metre}\times\SI{20}{\centi\metre}$ including penumbra.
    \item \textbf{Dimensions}: 
    The thickness of the outer housing shall be smaller than \SI{5}{\centi\metre} and fit in a cylinder of \SI{60}{\centi\metre} diameter. 
    The material shall correspond to a water equivalent thickness of less than 0.35 mm.
    \item \textbf{Environment}: 
    The detector must tolerate light, acoustic noise and magnetic fields of at least \SI{100}{\milli\tesla} in the fringe field of an MRI system (magnetic fields in the iso-center can reach up to \SI{1.5}{\tesla}).
\end{itemize}
The current beam monitoring system at HIT consists of two multi-wire chambers (MWC, see e.g.~\cite{sauli_PrinciplesOperationMultiwire_1977}) each with one plane for x- and y-direction to measure position and spot size and three ionization chambers (IC) to track the applied dose.
The MWC have a channel pitch of \SI{2}{\milli\meter}, but 
only a projection of the beam spot can be measured.
The signal from the first MWC is used to control a feedback loop for the beam position, the position is monitored by the second MWC. 
If the deviation between both measurements of position and width exceeds a threshold, a beam interlock is triggered.
Both chambers are also used to derive a spot size measurement, on which an interlock is defined, in case upper and lower thresholds are exceeded.
The integrated signal from the first IC controls the progress of the irradiation, and the beam is moved to the next raster point when the desired dose is reached. 
The other two ICs serve as independent checks of the particle fluence and ensure redundancy and diversity. 
If the deviation between the ICs exceeds a threshold, a beam interlock is triggered. \\
The MWCs, apart from their resolution, have some limitations which restrict their field of application. 
The concept of gas ionization and drift of charge carriers is limited to the drift velocity of the ions in gas. 
The signal is slightly delayed with respect to the first particles traversing the detector and there is still ion drift left when the particles have disappeared. 
These delays can be up to several hundreds of microseconds~\cite{lin_More10Years_2009}, which delays the interlocks, so that several raster points may be affected, before abort.
The performance of MWCs can also be affected by acoustic noise which can lead to vibrations of the wires and therefore deterioration of the resolution. 
One possible source of such acoustic noise can be an MRI system used for online position control of the target.
At the HIT QA room, an MRI system (Section~\ref{sec:Systems_MRI}) for developing MRI-guided ion beam therapy is available for testing the behavior of novel beam monitors.

\subsection{Alternative Beam Monitoring Concepts and Devices}
For more than 10 years, the community has been searching for improved detectors for high-rate beam monitoring. 
This section has no intention to be complete, but rather give a short introduction into some recent attempts.
An extensive review is given, for example, in~\cite{giordanengo_ReviewTechnologiesProcedures_2017}.\\
Some approaches are gas-based as the Time Projection Chamber-like approach described in~\cite{wang_BeamMonitorUsing_2017}, improved segmented ionization strip chambers with spatial resolution of \SI{100}{\micro\metre} in~\cite{actis_PreciseOnlinePosition_2014} or a more vibration tolerant version of an ionization chamber in~\cite{lin_More10Years_2009}. 
They all have in common that the charge collection time is in the order of milliseconds and very fast feedback on the timescale of few hundred microseconds is not possible.\\
One different approach is the measurement of the magnetic field of the deflecting magnets in the beam delivery system. 
The field strength is proportional to the final position of the beam spot and provide a very fast (\SI{10}{\micro\second}) and precise (\SI{10}{\micro\metre}) beam position measurement~\cite{klimpki_BeamMonitoringValidation_2017}. 
One drawback might be that the measurement is performed at quite a distance (more than about \SI{5}{\meter}) to the patient and the beam could be deflected by other sources, e.g. the magnetic field of an MRI, on its path.
In addition, remanence effects and dependencies on the beam position in the beam line can deteriorate the projected beam position at the target.\\
Scintillating fibers coupled to silicon photomultipliers (SiPM) are already used as fast, large area beam monitor~\cite{papa_ScintillatingFibresCoupled_2016, ortegaruiz_MultipurposeScintillatingFibre_2019a,  allegrini_CharacterizationBeamtaggingHodoscope_2021}. 
The drawback here is the weak radiation tolerance, which is limited to 1000 patient treatments for the device in~\cite{allegrini_CharacterizationBeamtaggingHodoscope_2021}. 
Additionally, a reduction of the attenuation length below 10\% after a dose of about \SI{50}{\kilo\gray} is indicated in~\cite{joram_LHCbScintillatingFibre_2015}.
Using photo diodes to read out scintillating fibers is investigated in~\cite{leverington_PrototypeScintillatingFibre_2018}. There, it is also suggested to use fibers with higher radiation hardness not only as natural attempt to overcome this weakness but also to profit from the shift in the emitted light spectrum.\\
An interesting attempt based on silicon is documented in~\cite{martisikova_TestAmorphousSilicon_2011}. 
The investigated large-scale detector ($\SI{20}{\centi\metre}\times\SI{20}{\centi\metre}$) is a commercial amorphous silicon based x-ray imager, but in this case applied to an ion beam. 
The resolution and stability over a period of four hours was very good, but the readout time was still high (\SI{80}{\milli\second\per frame}). 
The radiation tolerance over a period of one year needs to be evaluated.\\
Another promising silicon detector type are Ultra Fast Silicon Detectors (UFSD) based on thin Low Gain Avalanche Detectors (LGAD).
Three examples of applying this detector type in particle therapy applications can be found in~\cite{sacchi_TestInnovativeSilicon_2020, vignati_ThinLowgainAvalanche_2020, kruger_LGADTechnologyHADES_2022}. 
These detectors can be used for very fast timing applications like time-of-flight detector systems. 
The good time resolution of better than \SI{100}{\ps} is a result of the intrinsic signal gain of these devices, which is about 10 to 20.
This gain (and the time resolution) is reduced significantly after irradiation. 
A gain reduction of 20\% after a fluence of \SI{1e12}{\per\square\centi\metre} is quoted in~\cite{sacchi_TestInnovativeSilicon_2020}.
This is small compared to the yearly expected fluence in the center of a beam monitoring detector. 
This detector type requires external readout electronics. 
For a pixellated layout, the readout electronics needs to be connected via bump bonds to the LGAD, adding material and complexity during the assembly.\\
To avoid the interconnection step, reduce material and bring the first signal amplification stage close to the generated signal, monolithic detector chips have been developed. 
In~\cite{flynn_MonitoringPencilBeam_2022} a commercial large scale CMOS detector for x-ray imaging was evaluated as proton beam monitor. 
The determination of the beam profile was demonstrated to reproduce measurements with films with minor deviations as long as saturation is not reached. 
The rolling shutter readout mode of this device limits the readout speed and a full readout of the $\SI{12}{\centi\metre}\times\SI{14}{\centi\metre}$ large matrix leads to saturation effects already at low beam currents. 
For this device, no evaluation of the radiation tolerance was reported in~\cite{flynn_MonitoringPencilBeam_2022}.

\subsection{A Beam Monitor Based on HV-CMOS Detectors} \label{sec:proposed_system}
In this article, a novel approach is discussed based on custom HV-CMOS detectors, introduced in Section~\ref{sec:HV-CMOS}.
These monolithic devices combine the sensing element with signal processing in one thin silicon chip. 
For fast readout, the chips implement particle counting and projection of particle counts on the x- and y-axis.
The reticle-sized chips will be arranged in a $13 \times 13$ array to achieve the required active area of $\SI{25}{\centi\metre}\times\SI{25}{\centi\metre}$ and interconnected by thin flex PCBs to the surrounding electronics, where electrical signals are converted to optical signals which allow stable data transmission to a fast field-programmable gate array (FPGA) board for further data processing.
The chip array is mounted on thin carbon fiber plates for mechanical and thermally dissipative support.
\begin{figure}
    \centering
    \includegraphics[width=0.8\textwidth]{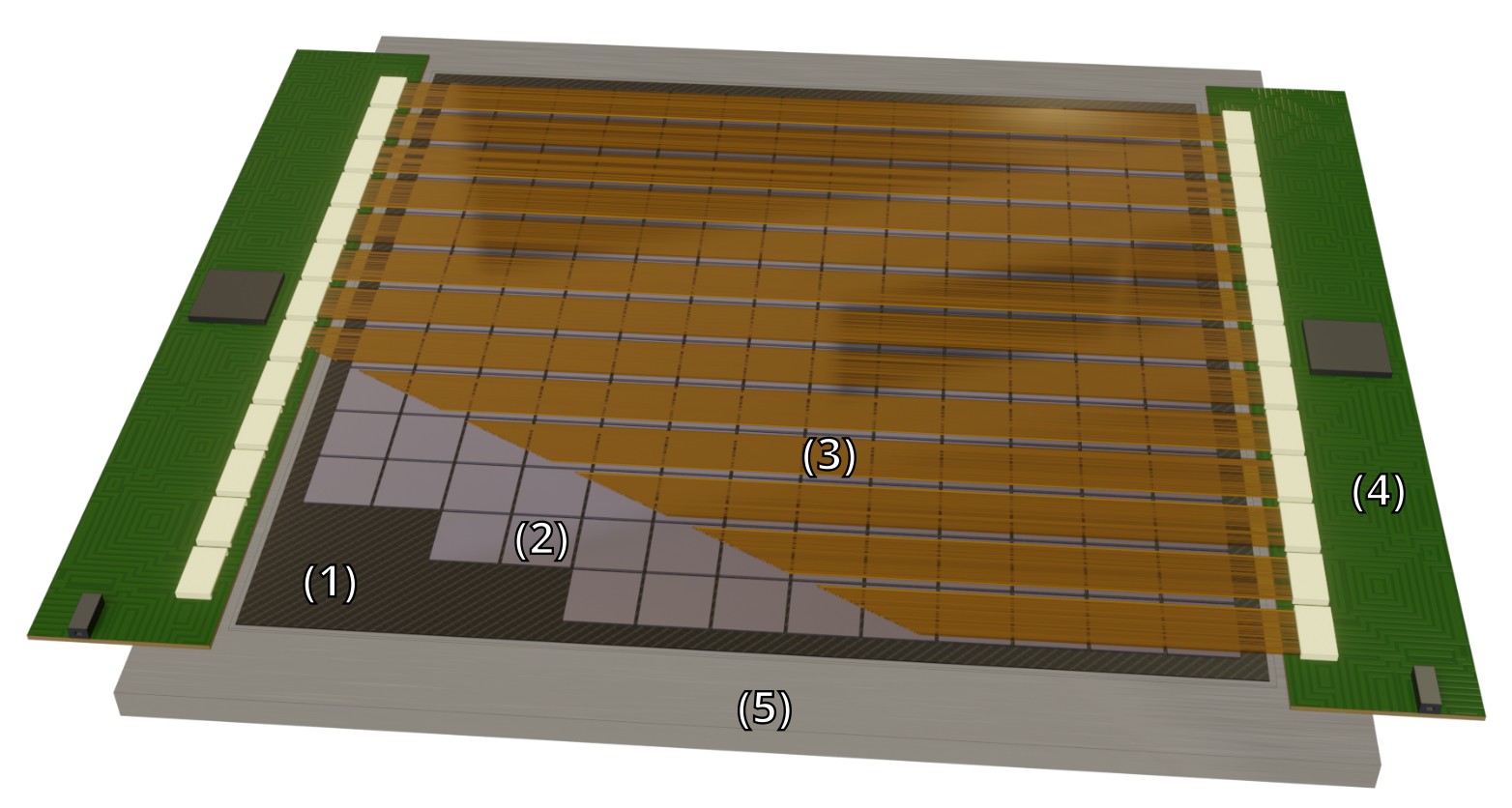}
    \caption{Schematic 3D rendering of the full size matrix with $13 \times 13$ chips (2), interconnected by flex cables (3), assembled on a thin carbon fiber plate (1). The surrounding electronic boards (4) are mounted on a stabilizing frame (5).}
    \label{fig:FullMatrixDrawing}
\end{figure}
Figure~\ref{fig:FullMatrixDrawing} shows a schematic 3D rendering of the final detector design.\\
After the introduction of the chip technology in Section~\ref{sec:HV-CMOS}, the systems and facilities used for the characterizing measurements are described in Section~\ref{sec:Systems} and, finally, the evaluation of the detector and system performance is discussed in Section~\ref{sec:Studies}.
%%%%%%%%%%%%%%%%%%%%%%%%%%%%%%%%%%%%%%%%%%
\section{A Beam Monitoring Chip based on HV-CMOS Technology}
\label{sec:HV-CMOS}
\subsection{Silicon Sensor Technology}
Particle detectors made of silicon base material are widely used in experimental particle physics~\cite{hartmann_SemiconductorSensors_2011} and proposed for usage in radiotherapy for many years~\cite{bruzzi_NovelSiliconDevices_2016}. 
They are operated in high magnetic fields (up to \SI{3.8}{\tesla}) and high particle fluxes (up to \SI{3.5}{\giga\hertz\per\square\centi\metre}). 
The basic building blocks are pn-junctions operated in reverse bias. 
Traversing charged particles generate electron-hole pairs by ionization (about 75 per \si{\um}) in the depleted silicon bulk and the readout electronics detects the produced current pulses. 
For hybrid systems, the readout electronics is implemented on dedicated ASICs which are connected to the silicon sensor by wire bonds in case of planar strip sensors or bump bonds for pixelated sensors~\cite{hartmann_EvolutionSiliconSensor_2017, lutz_SemiconductorRadiationDetectors_2007}.
The hybrid approach allows to combine a specialized sensor with specialized readout electronics, which are developed independently. 
Still, the hybridization requires additional and potentially expensive production steps.
It also limits the granularity and mass (saving) of the final detector.
An alternative are monolithic devices described in the next section.

\subsection{The HV-CMOS Technology}
CMOS technology allows to combine the sensitive pn-junction with the readout electronics on one thin (less than \SI{150}{\micro\metre}) chip.
While keeping the high granularity of a pixel sensor, this approach avoids the additional readout chip (reducing material thickness) and the connectivity step (reducing costs).
The radiation environment in experimental particle physics and beam monitoring calls for silicon chips with increased radiation tolerance.
One possibility is to use a commercially available High-Voltage CMOS (HV-CMOS) process~\cite{peric_ParticlePixelDetectors_2011}.
This technology allows to apply a bias voltage of up to about \SI{300}{\volt} generating a wide depletion zone (typically \SIrange{35}{50}{\micro\metre}).
Generated charge carriers are then collected by fast drift (order of few \si{\ns}) in the electric field and not only by slow diffusion as in standard CMOS devices. 
The faster drift reduces the effect of charge carrier trapping~\cite{moll_RadiationDamageSilicon_1999} in irradiated silicon and more charge carriers are generated in the wider depletion zone yielding a higher signal.
At least the electronics of the first steps of signal processing is located inside the pixels (smart diode).
Typically, also signal amplification and digitization circuits are added.
Combining improved radiation tolerance with integrated electronics and small pixel sizes makes the HV-CMOS technology a promising candidate for a beam monitoring system.
Several HV-CMOS based detectors have been developed for particle physics experiments so far~\cite{arndt_TechnicalDesignPhase_2021, augustin_MuPixSensorMu3e_2020,  schimassek_TestResultsATLASPIX3_2021, peric_HighVoltageCMOSActive_2021}.

\subsection{An HV-CMOS Detector for Beam Monitoring}\label{sec:det_for_BM}
The beam monitoring detector chip HitPix has been developed in \SI{180}{\nano\metre} HV-CMOS technology and was produced in two generations.
HitPix1 has $24 \times 24$ pixels, is $\SI{5}{\milli\metre}\times\SI{5}{\milli\metre}$ in size and is described in detail in~\cite{weber_HighVoltageCMOS_2022, weber_DevelopmentIntegratedCircuits_2021}.
HitPix2 has $48 \times 48$ pixels, is $\SI{10}{\milli\metre}\times\SI{10}{\milli\metre}$ in size and is the generation investigated in this article.
The dimensions of a detector chip for the envisaged final beam monitoring system will be $\SI{20}{\milli\metre}\times\SI{20}{\milli\metre}$ with $96 \times 96$ pixels.
In this section, only the main features of the HitPix family in view of the targeted application are summarized.\\
The very high particle rates can be processed by counting the particle transitions in each pixel cell during a programmable period of several microseconds. 
One hit is registered when the induced signal of a charged particle traversing a pixel cell passes a programmable threshold, which needs to be optimized with respect to noise and induced signal.
The hit counts are accumulated over a programmable time period.
The resulting count status of the pixel matrix is called \textit{frame}.
One frame can then be read out, either for each pixel of the matrix (\emph{counter readout}, slow) or for the sums of each row and column (\emph{adder readout}, fast), which are calculated on-chip. 
In the counter readout mode the readout takes longer since more data (\SI{129024}{bits} per frame for the final chip) is handled.
It is used for debugging or detailed beam diagnostics, since some skewed beam shapes can only be detected with two-dimensional measurements.
In the adder readout mode, the projected data (\SI{2688}{bits} per frame for the final chip) can be read out 48 times faster, while the pixel hit counting continues.
The adding step happens quasi instantaneous when any counter in the corresponding row/column is increased, and all sums are transferred to the shift register by one load pulse when the readout is triggered.
\begin{figure}
    \centering
    \begin{subfigure}[b]{0.3\textwidth}
        \includegraphics[width=\textwidth]{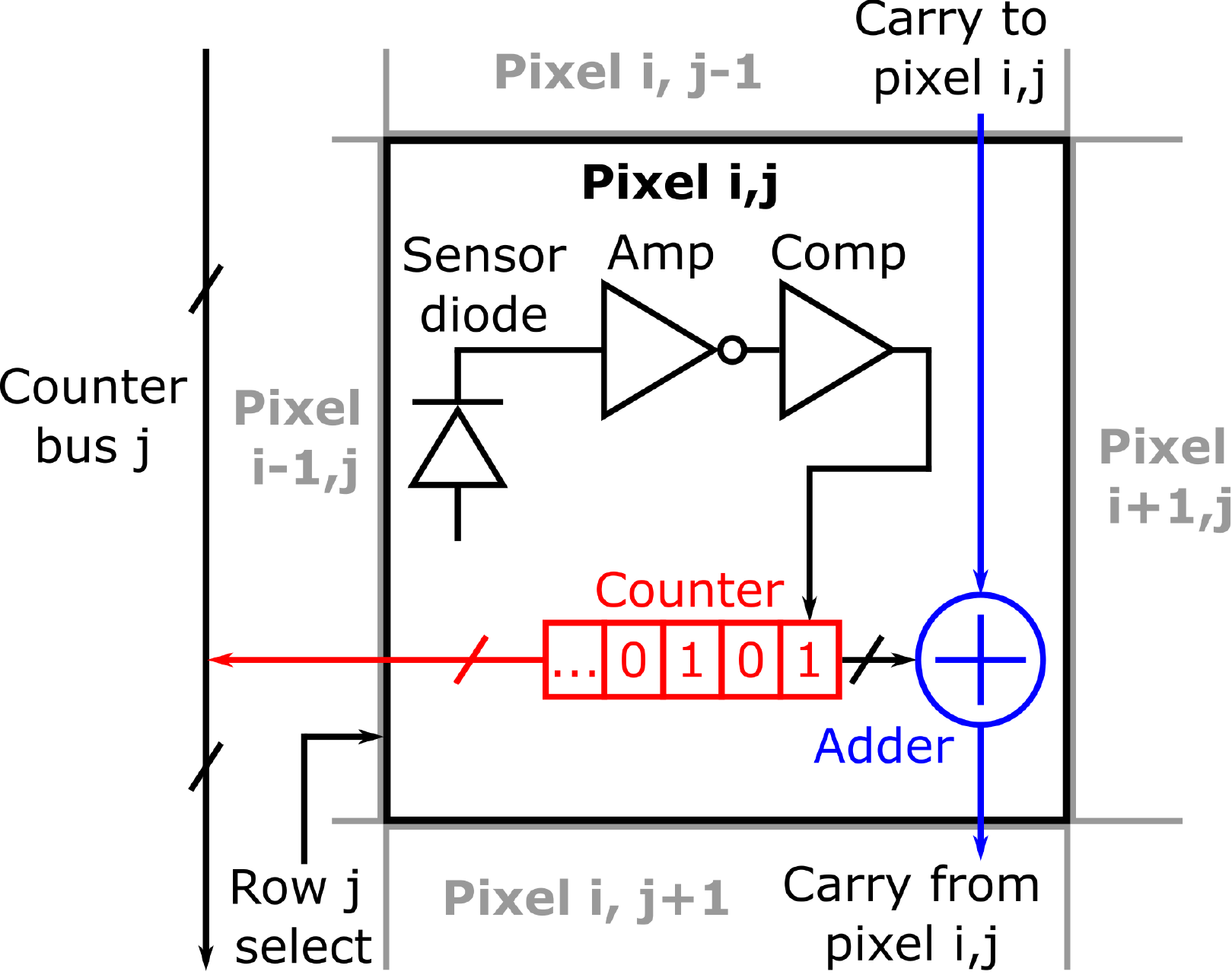}
        \caption{Components in the pixel cell}
    \end{subfigure}
    \hfill
    \begin{subfigure}[b]{0.3\textwidth}
        \includegraphics[width=\textwidth]{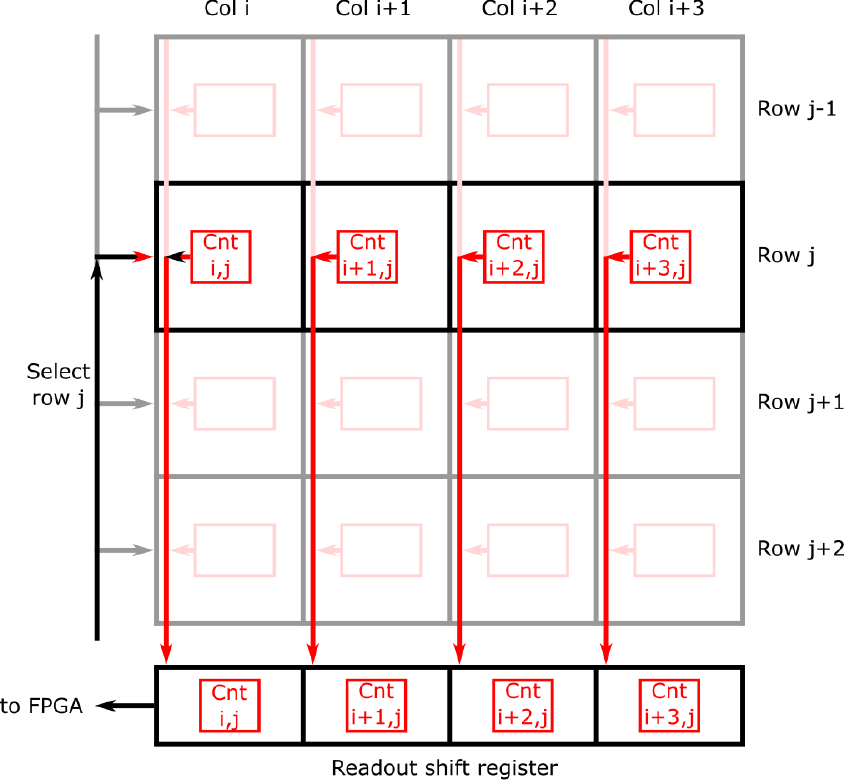}
        \caption{Pixel matrix readout in counter mode}
    \end{subfigure}
    \hfill
    \begin{subfigure}[b]{0.3\textwidth}
        \includegraphics[width=\textwidth]{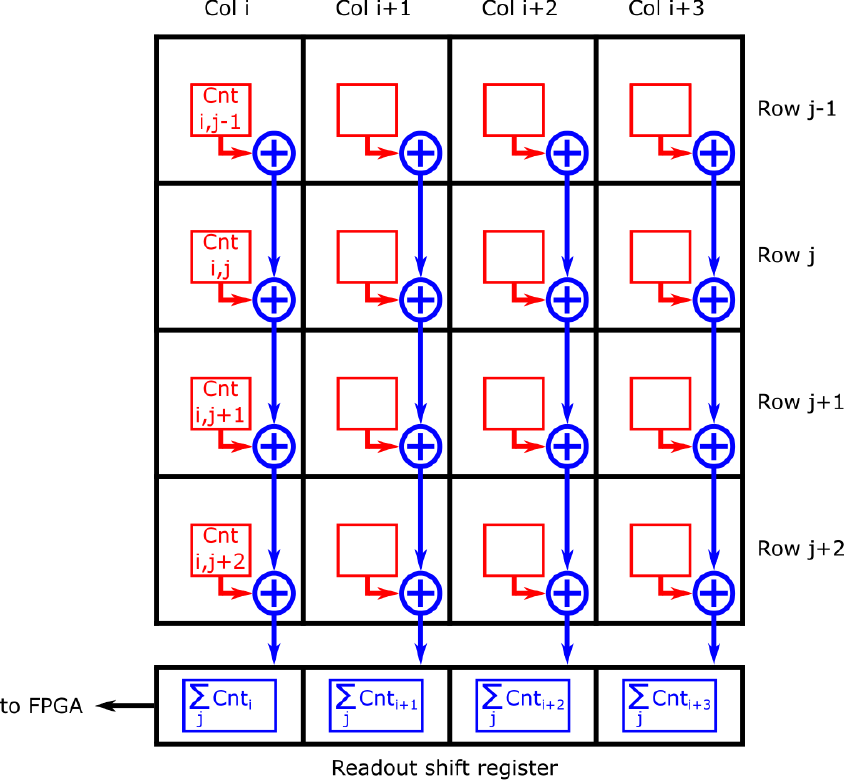}
        \caption{Pixel matrix readout in adder mode}
    \end{subfigure}
    \caption{Illustration of the hit counting and adding features of the HitPix. In (a), the signal from the sensitive diode (Sensor diode) is amplified (Amp) and a threshold is applied in the comparator (Comp). The hits are counted and stored in the counter. This value can either be read out via the counter bus (left path in (a), and (b)), or added to the sum being propagated from the preceding pixel and then passed on to the next pixel (blue path in (a) , and (c)).}
    \label{fig:HitPix_Pixel}
\end{figure}
The concept is illustrated in Figure~\ref{fig:HitPix_Pixel}.\\
It is clear that the projections in adder readout mode do not contain the full information on the beam shape, but the most relevant like position and width.
The shape is typically constant on the scale of an irradiation day, therefore the projections of the beam profile are sufficient to generate a fast interlock in case the beam behaves unexpectedly.\\
The required pixel size was evaluated from cluster size measurements with available HV-CMOS detectors featuring a similar charge collecting diode structure, but smaller pixels, and the requirement to reach at least \SI{200}{\micro\metre} position resolution.
The ionization of therapeutic ions is much larger than minimum ionizing particles as they typically appear in particle physics experiments. 
Protons in the range given in Table~\ref{tab:beampara} generate two to six times more charge and carbon ions 50 to 135 times more.
Although the deposited charge is much higher, the average cluster size was measured to be below two pixels for $\SI{150}{\micro\metre}\times\SI{50}{\micro\metre}$ pixels for these type of ions~\cite{schimassek_EntwicklungUndCharakterisierung_2021}.
This concludes that the charge of one particle can be well contained in a $\SI{200}{\micro\metre}\times\SI{200}{\micro\metre}$ pixel also fulfilling the requirement for a resolution of at least \SI{200}{\micro\metre}.\\
Another aspect for the chip design is a narrow inactive periphery, which was achieved by locating data processing (analog-to-digital converters, hit counters, hit adders) and data storage inside each pixel.
The periphery houses only configuration storage and a simple shift register readout.\\
The HitPix2 is prepared for daisy-chain readout of several detectors to minimize readout lines on the interconnecting cable, when moving to a multi-chip array.
The chain can be up to seven detectors long.
For the final system, the detectors will have $96 \times 96$ pixels and two projections with \SI{14}{bits} per line.
This results in \SI{18816}{bits} for a chain containing seven detectors to be read out in adder mode during the integration time of one frame, which is in the order of \SI{20}{\micro\second} depending on the beam intensity.
To achieve this, a data rate of about \SI{1}{\giga\hertz} is required.
In case the readout time for many detectors in a daisy-chain turns out to not meet the specifications, each detector has an additional direct readout option. 
In this mode the selection of individual detectors for readout is possible to define a region-of-interest for each frame and, to save data traces on the flex cable, allows using a common data bus structure sharing the signals of some detectors on one line.

%%%%%%%%%%%%%%%%%%%%%%%%%%%%%%%%%%%%%%%%%%
\section{Irradiation Facility and Test Systems}\label{sec:Systems}
\subsection{Irradiation facility}\label{sec:irradiation_facilities}
To accumulate a significant particle fluence in short time, a compact cyclotron with \SI{23}{\mega\electronvolt} protons was used.
The cyclotron is operated by the company ZAG Zyklotron AG\footnote{\url{https://www.zyklotron-ag.de/index.php}} (ZAG) and situated on the Campus North of KIT. 
The beam has a width of about \SI{8}{\milli\metre} (FWHM) and the beam current is set to about \SI{2}{\micro\ampere} for irradiation of the detectors.
At this beam current, the samples are irradiated to the target fluence of around \SI{1e15}{\per\square\centi\metre} within \SI{15}{\minute}.
The samples are placed inside an insulated box which is flushed with cold nitrogen gas (below \SI{-20}{\degreeCelsius}). 
The box is scanned in front of the beam to achieve a uniform irradiation of the samples.
The damage to the silicon lattice caused by these protons can be converted into an equivalent damage caused by \SI{1}{\mega\electronvolt} neutrons by the non-ionization energy loss (NIEL) hypothesis~\cite{lindstrom_RadiationHardnessSilicon_1999}. The particle fluence $F_\mathrm{particle}$ can be converted to the equivalent fluence as $F_\mathrm{eq} = \kappa \times F_\mathrm{particle}$.
The corresponding scaling factor is called hardness factor $\kappa$, with $\kappa=2.0$ for \SI{23}{\mega\electronvolt} protons.

\subsection{Magnetic Field Test Environments}\label{sec:Systems_MRI}
\begin{figure}
    \centering
    \begin{subfigure}[b]{0.45\textwidth}
        \includegraphics[height=6cm]{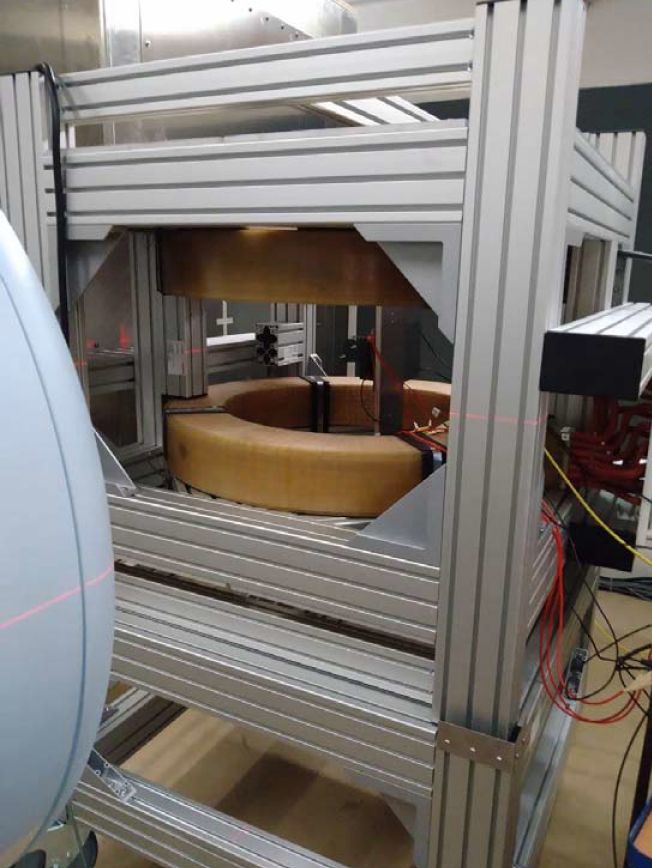}
        \caption{Helmholtz coil pair} \label{fig:HHC}
    \end{subfigure}
    \quad
    \begin{subfigure}[b]{0.45\textwidth}
        \includegraphics[height=6cm]{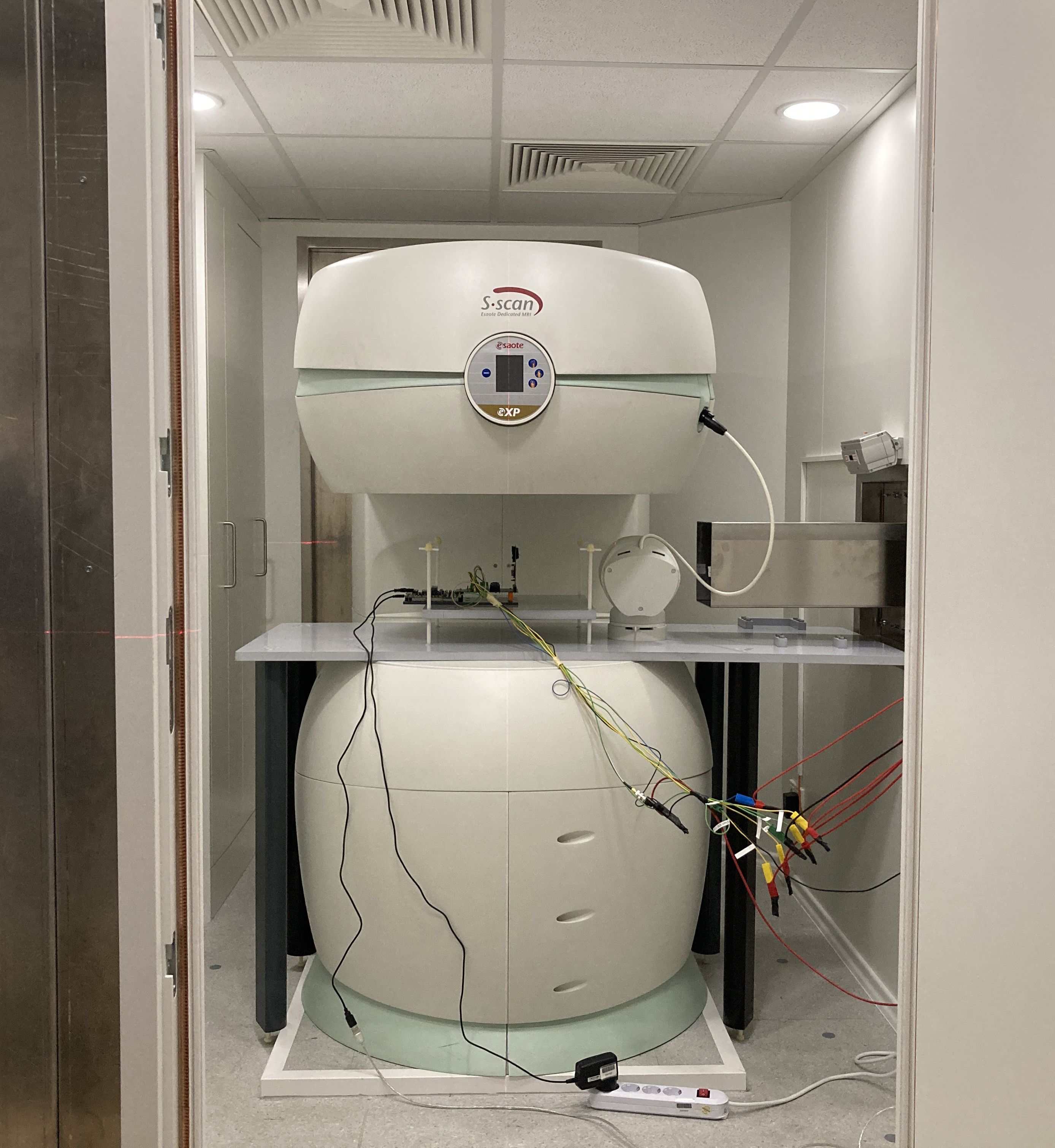}
        \caption{MRI system} \label{fig:MRI}
    \end{subfigure}
    \caption{Pictures of the Helmholtz coil pair (a) and MRI system (b). Both are used for measurements at HIT. The Helmholtz coils allow in-beam characterization of devices up to a static magnetic field of \SI{100}{\milli\tesla}. In the MRI system, the ion beam enters from the right side through the wave guide. The whole setup is located inside a movable enclosure for shielding the stray fields. Image source: HIT.}
    \label{fig:B-setups}
\end{figure}

The HV-CMOS technology does not only have the potential to improve the performance of the existing beam monitoring system, but the tolerance of silicon sensors towards magnetic fields promises the simultaneous irradiation and MR imaging of the target.
While most sensor technologies can be operated in a certain distance to the high field region of an MRI system, HV-CMOS sensors are expected to work within the center of the field.
HIT has built a large scale Helmholtz coil pair with an inner diameter of \SI{60}{\centi\metre} for measurements in a magnetic field of up to \SI{100}{\milli\tesla} under lab condition.
A picture of the setup is shown in Figure~\ref{fig:HHC}.
It was used for basic measurements in a static field with the HV-CMOS detectors.\\
In order to measure the effect on the detector of the even stronger field of an MRI magnet and its alternating field during scanning, an MRI system\footnote{Esaote S-Scan compact} was used which is available in the QA room at HIT with a static magnetic field of \SI{250}{\milli\tesla}.
The MRI system is located inside an enclosure for shielding the stray fields as shown in Figure~\ref{fig:MRI}.
The whole setup can be moved into the beam for experiments.
It was used for the tests reported in Section~\ref{sec:Studies_MRI}.

\subsection{Data Acquisition System}
\begin{figure}
    \centering
    \includegraphics[width=0.95\textwidth]{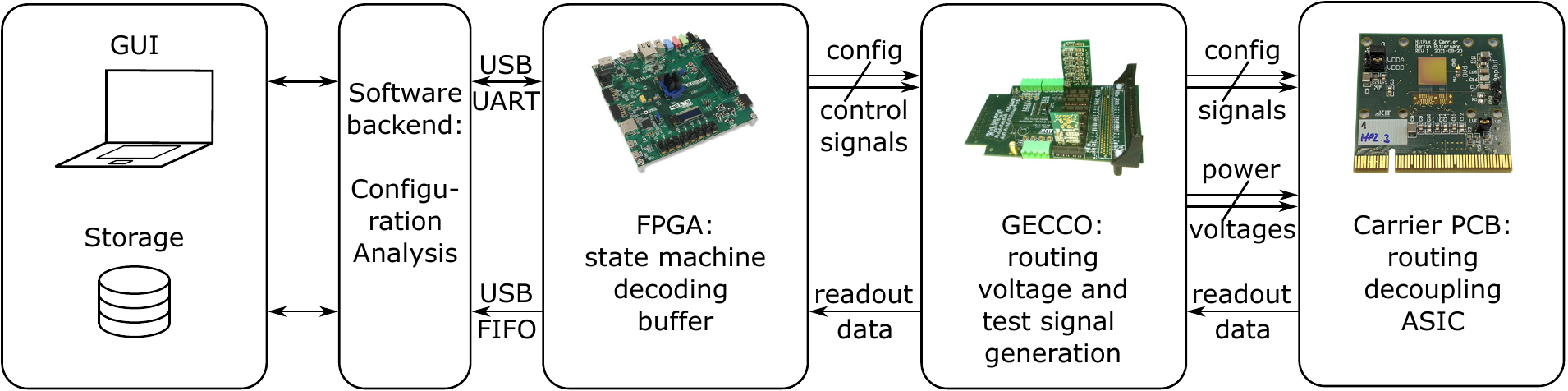}
    \caption{Block diagram of the data acquisition chain with the four components: PC, FPGA board, GECCO board and carrier board for the detector. Communication and data paths are indicated between the blocks.}
    \label{fig:DAQ}
\end{figure}

The test system to control and read out the detectors is split into four parts.\\ 
The detectors are mounted and wire-bonded on a simple PCB called \emph{carrier}. 
The carrier can be plugged into a PCIe socket of the GEneric Configuration and COntrol board (GECCO)~\cite{ehrler_CharakterisierungMonolithischenHVCMOSPixelsensoren_2021, schimassek_EntwicklungUndCharakterisierung_2021}. 
GECCO provides all necessary supply voltages, test signals and routes data lines from and to an FPGA.
In this case, a Nexys Video Artix-7 FPGA board\footnote{\url{https://digilent.com/shop/nexys-video-artix-7-fpga-trainer-board-for-multimedia-applications/}} is used.\\
The firmware to control the detector implements a state machine. 
Communication to the HitPix detector is established via a two phase shift register driven by the FPGA with up to \SI{200}{\mega bit\per\second}.\\
Finally, the hardware is connected to a PC by USB. 
The software interacting with the FPGA uses multi-threading to receive and decode data from the FPGA as quickly as possible while online data processing can continue.\\
Figure~\ref{fig:DAQ} shows a block diagram of the data acquisition chain with the different elements and the data streams.
%%%%%%%%%%%%%%%%%%%%%%%%%%%%%%%%%%%%%%%%%%
\section{Detector Characterization and Evaluation as a Beam Monitoring Device}
\label{sec:Studies}
This section summarizes the characterization of the detector technology and describes the evaluating studies in view of the beam monitoring application performed so far. 

\subsection{Technology Decisions}
\begin{figure}
    \centering
    \begin{subfigure}[b]{0.45\textwidth}
        \includegraphics[width=\textwidth]{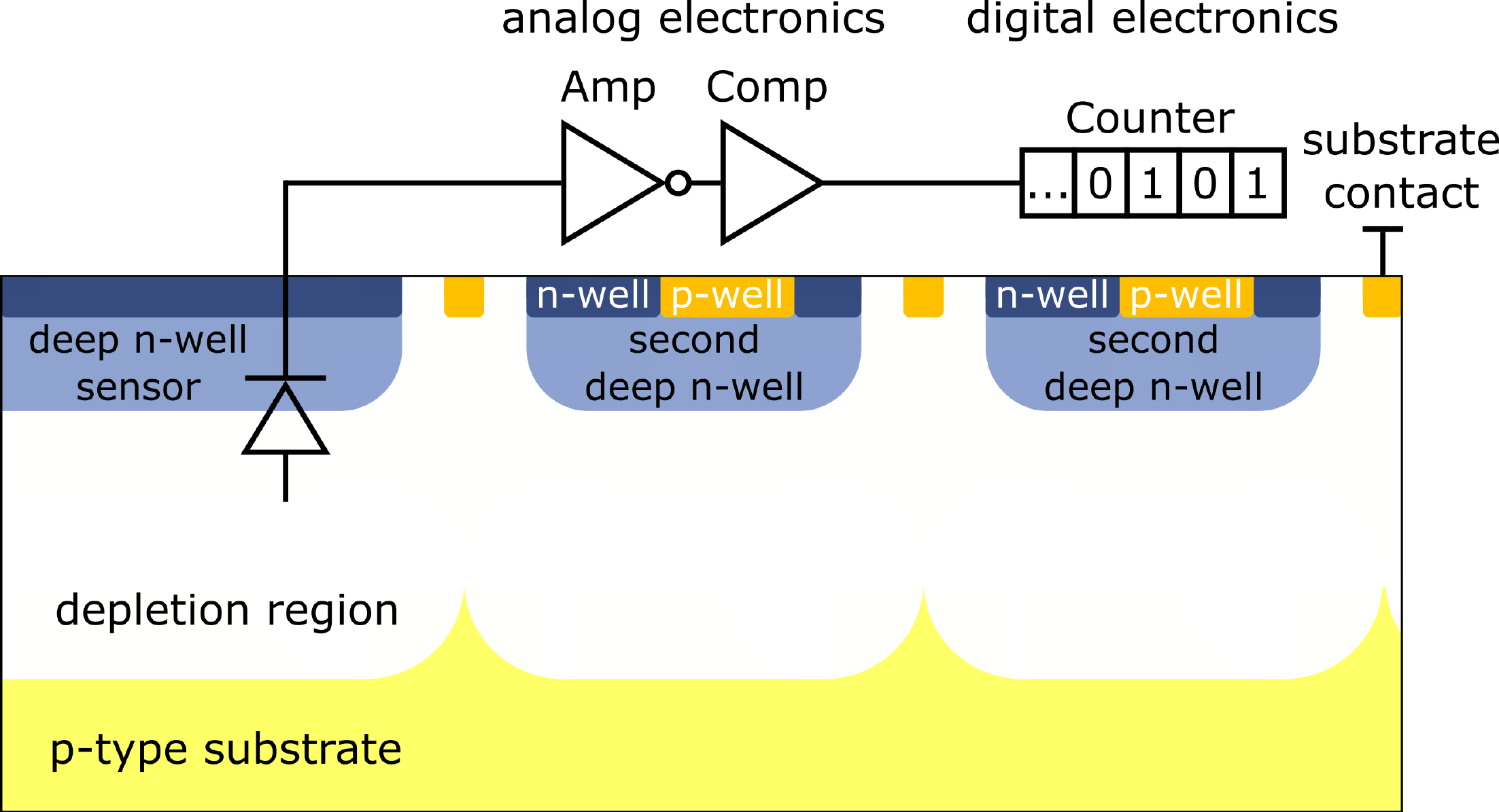}
        \caption{Standard cell with separated n-wells (STD)\\}
    \end{subfigure}\hfill
    \begin{subfigure}[b]{0.45\textwidth}
        \includegraphics[width=\textwidth]{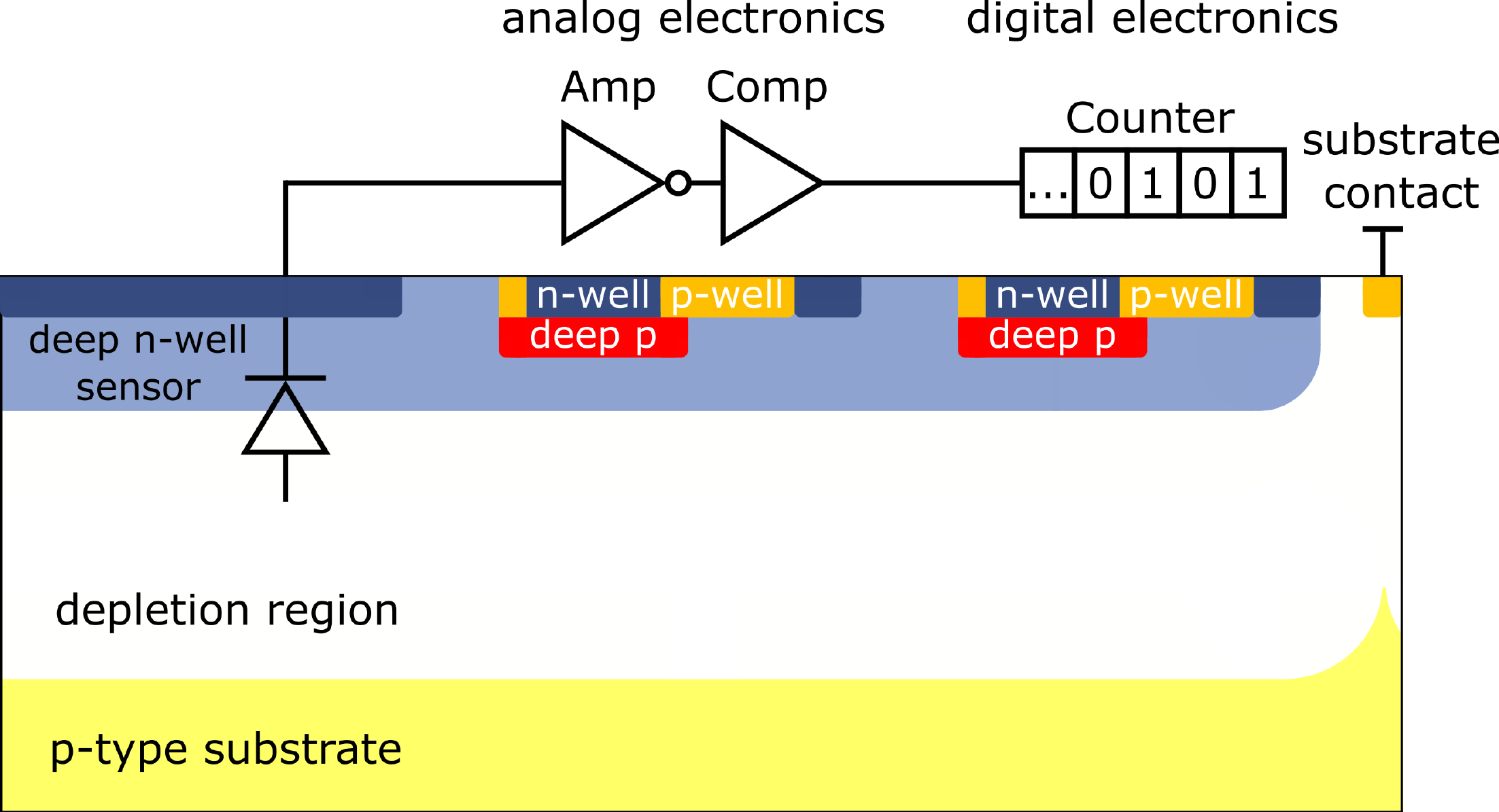}
        \caption{Common n-well design with isolating deep p-well (ISO)}
    \end{subfigure}
    \caption{Pixel cross sections of the two designs. In the STD design the n-wells are separated for the charge collecting diodes and electronics (a). A common n-well enlarges the collecting diode, but requires deep p-wells to isolate the electronics (b). After~\cite{weber_HighVoltageCMOS_2022}.}
    \label{fig:iso}
\end{figure}

\begin{figure}
    \centering
    \begin{subfigure}[b]{0.45\textwidth}
        \includegraphics[width=\textwidth]{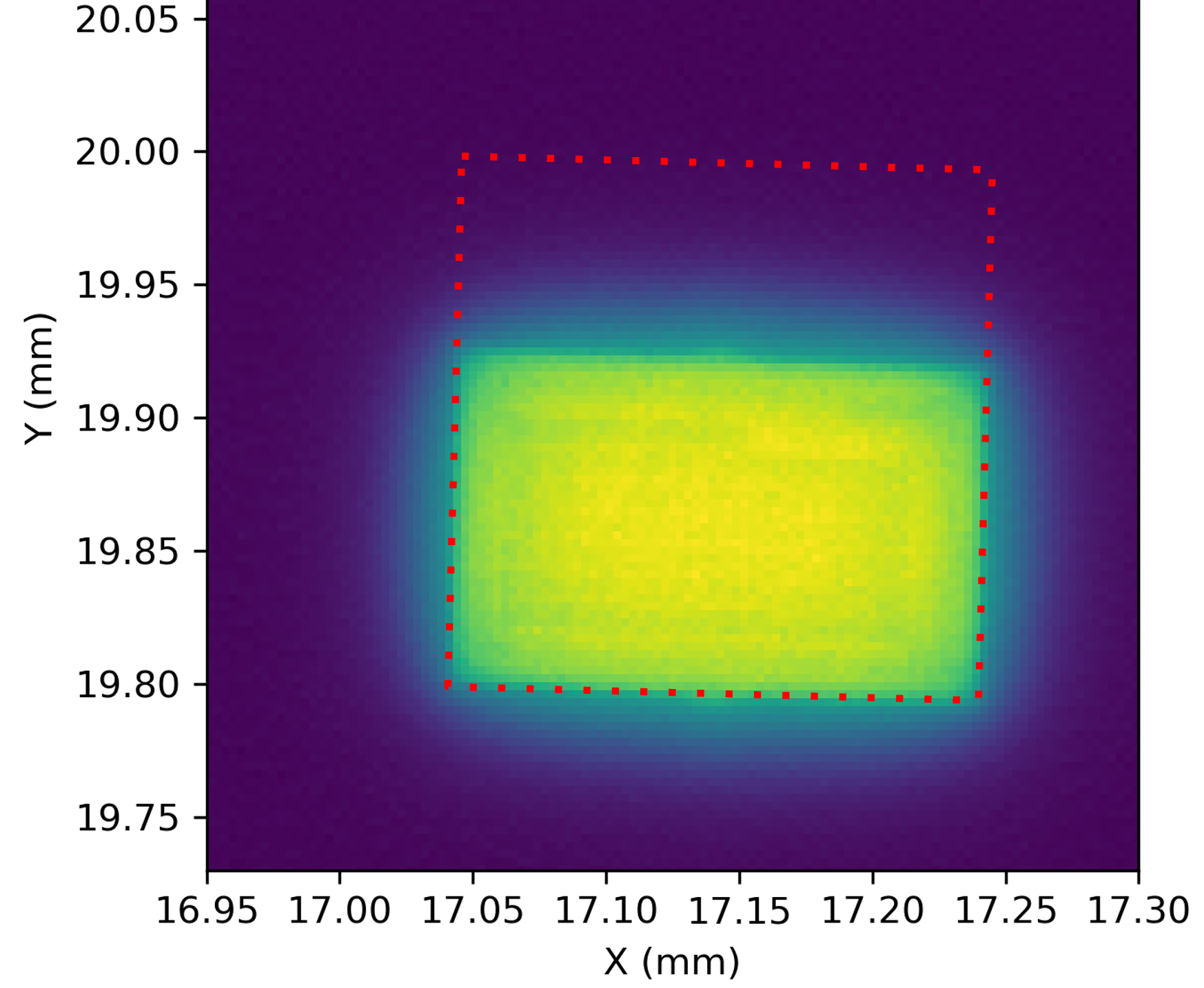}
        \caption{Signal height map for the STD design}
    \end{subfigure}\hfill
    \begin{subfigure}[b]{0.45\textwidth}
        \includegraphics[width=\textwidth]{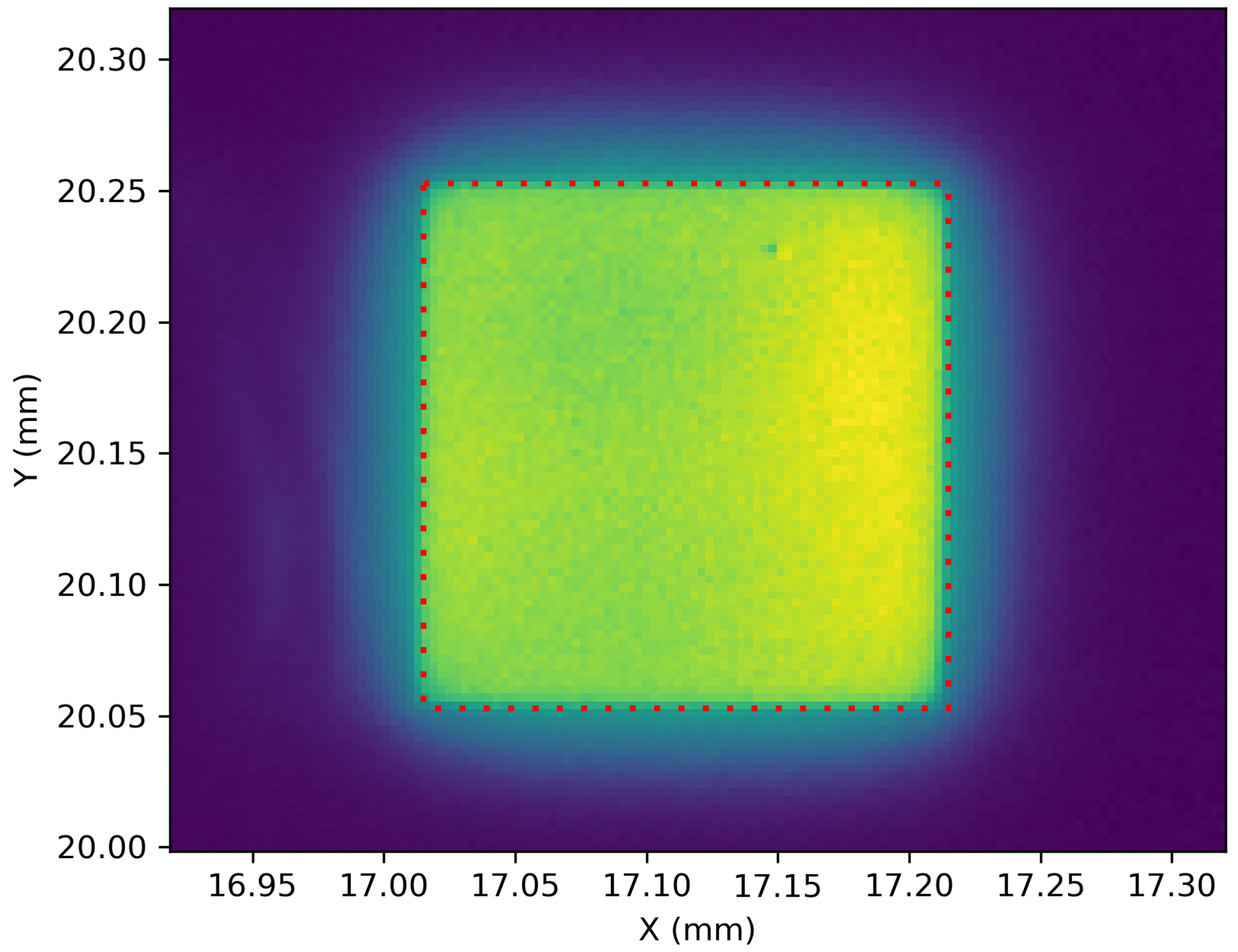}
        \caption{Signal height map for the ISO design}
    \end{subfigure}
    \caption{Signal height distribution measured on one pixel cell during illumination of the detectors with an infrared (\SI{904}{\nm}) laser from the backside for the STD cell (a) and the ISO cell (b). The laser spot size was about \SI{10}{\micro\metre} (FWHM) while the spatial step size was \SI{1}{\micro\metre}. The setup is described in more detail in~\cite{schimassek_EntwicklungUndCharakterisierung_2021}. The analog signal of the pixel can be read out on a dedicated output line.}
    \label{fig:illumination}
\end{figure}

Since the beam monitoring detector should register nearly all traversing particles, the insensitive regions need to be very small. 
Therefore, the insensitive periphery of the detector can only occupy a small fraction of the overall detector area. 
This is possible by using simple and area saving building blocks for the communication part and moving data processing blocks into the pixel cells.\\
The first version of HitPix has been produced in two flavors.
One is designed in standard technology (STD) with separated charge collecting diode and deep n-well housing the electronics. 
The other flavor (ISO design) has additional deep p-wells isolating the shallow n-wells from the common deep n-well allowing for the collection diode to cover the whole pixel area.
This difference is visualized in Figure~\ref{fig:iso}.
Illuminating the pixel cells with an infrared laser from the backside confirmed that the area below the electronics part of the standard design is insensitive to the generated charge, while the design with the isolating deep p-well is sensitive in the entire pixel cell. 
Figure~\ref{fig:illumination} shows that about one third of the STD pixel cell area is insensitive in this measurement.
This effect was expected, since this is the fraction of the pixel cell occupied by the electronics part. 
However, the clusters by protons and carbon ions are sufficiently large so that even a particle passing in the middle of the insensitive area will still leave a signal in the adjacent pixel diodes.
Even if this was not the case, the inefficiency would be predictable and on average the lost signal could be corrected for.\\
Another effect of the common deep n-well in the ISO design is an increase in detector capacitance, leading to smaller and shorter induced signals, which could lead to a critically low signal-to-noise ratio with minimum ionizing particles.
The signals from an ion beam, however, are comparably large (Section~\ref{sec:det_for_BM}) and could provoke extended dead times while the signals exceed the threshold.
In this case, shorter signals would be beneficial to increase the rate capability.\\
The radiation tolerance of the detector needs to be addressed already at the design stage. 
Therefore, the design of HitPix2 uses p-MOS electronics where possible and enclosed n-MOS transistors where needed.
Besides being more radiation tolerant~\cite{snoeys_LayoutTechniquesEnhance_2000}, an amplifier built solely with n-MOS transistors requires only one supply voltage.
This reduces the number of power supply pads and, therefore, inactive area in the periphery.\\
The depth of the depleted zone and the electric field below the collecting diode depends on the substrate resistivity and bias voltage, and has a significant effect on the induced signal.
Three different resistivities of the silicon substrate have been tested with HitPix2 and the results are discussed in Section~\ref{sec:radiation_tolerance}.\\
The chip is designed for low power consumption to minimize the challenges of cooling the matrix in this low mass application. 
The total power consumption measured on HitPix2 is \SI{20}{\micro\watt} per pixel or \SI{47}{\milli\watt\per\square\centi\meter}, split into equal fractions for analog and digital power.
This is comparable to other low power CMOS detectors like the ALPIDE chip with \SI{40}{\milli\watt\per\square\centi\meter}~\cite{mager_ALPIDEMonolithicActive_2016}, which was designed for much lower particle density though.

\subsection{Hit Detection and Noise}
\label{sec:hit_detection}
\begin{figure}
    \centering
    \includegraphics[width=0.75\textwidth]{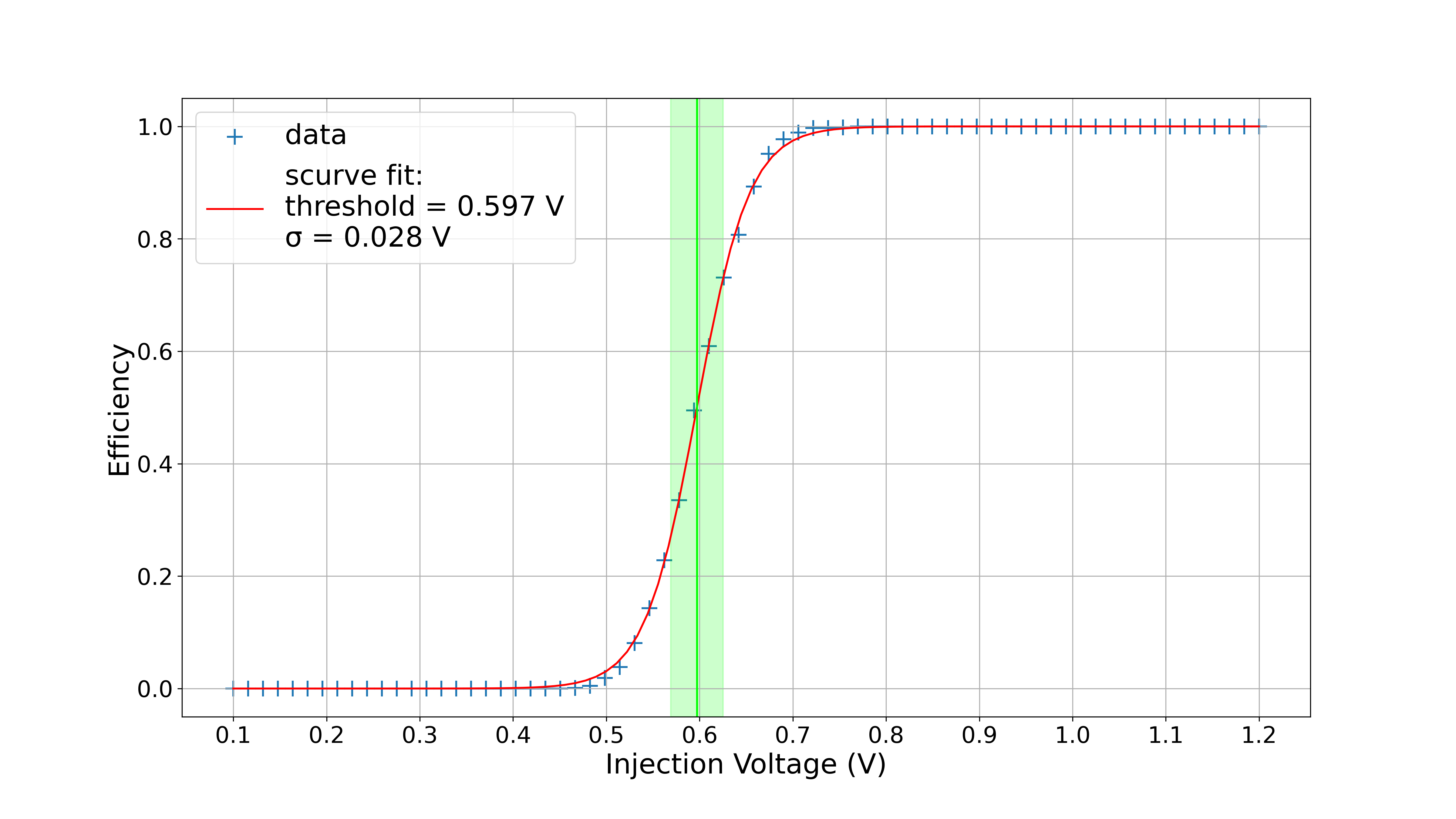}
    \caption{Typical response curve of a pixel to injected charge pulses. Here, efficiency is defined as the ratio of detected hits to injected pulses. The data can be fit by an error-function as indicated by the red curve. The green line and band mark the threshold voltage and the one standard deviation noise amplitude. The noise figure here is units of voltage in the injection circuit and not to be compared with noise voltage at the amplifier output.}
    \label{fig:scurve}
\end{figure}

\begin{figure}
    \centering
    \includegraphics{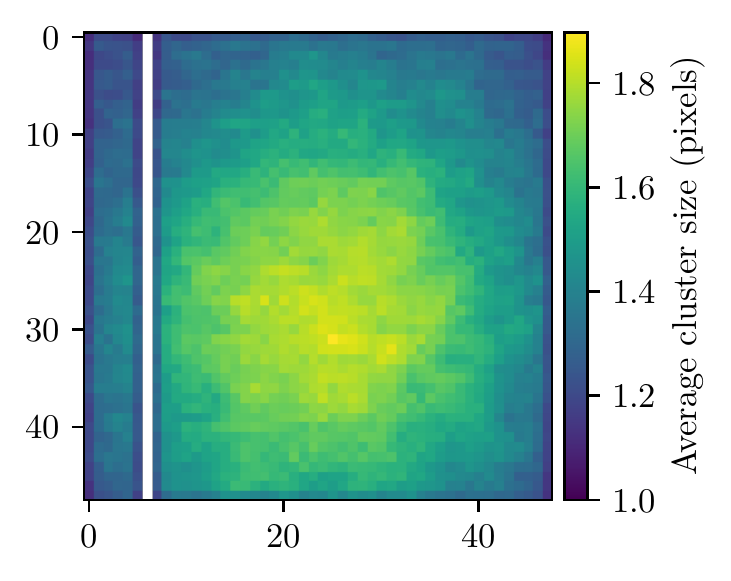}
    \caption{Average cluster size over the pixel matrix for a carbon ion beam from~\cite{pittermann_EvaluationHVCMOSSensors_2022}. The higher cluster size in the center corresponds to pileup effects (described in the text). The white line is caused by a defective column which was masked for this analysis. Beam parameters: carbon ions, energy range of \SIrange{108}{423}{\mega\electronvolt\per u}, \SI{5e6}{\per\second} particle rate.}
    \label{fig:clusters_run_030}
\end{figure}

Ionizing particles are registered by the HitPix detector when the induced signal exceeds a programmable threshold.
The signal height is not measured by the standard data processing chain.
For debugging purpose, the analog signals of one row can be routed to a readout pad.
To determine the noise of the pixel cells and to understand the response to electrical test signals, an external voltage pulse of varying amplitude can be injected into the pixel front-end.
By scanning the amplitude of the injected signal at a fixed threshold, the pixel response and noise can be determined.
The distribution follows an error-function and the noise corresponds to the width of the transition region.
Figure~\ref{fig:scurve} shows an example of such a measurement.
The general noise level of HitPix2 at the comparator stage is about \SI{28}{\milli\volt}, while a typical signal from high energy protons is about \SI{200}{\milli\volt}. 
Knowing the noise level and the expected signal height of the relevant particles, one can set an optimal threshold, which eliminates hit counts due to noise, but still registers all particle transitions.\\
To verify that the chosen pixel size is adequate for the ion beam, the cluster size was evaluated for different beam conditions.
A cluster is a group of adjacent hits.
If the cluster size is close to one, the generated charge carriers are mainly localized within one pixel, which is advantageous for this counting application.
Figure~\ref{fig:clusters_run_030} shows the average cluster size distribution in a carbon ion beam with small spot size.
The larger cluster sizes in the center are mainly from pileup.
That means that within the integration time of one \SI{50}{\micro\second} frame many hits are registered and the probability increases that two hits appear in adjacent pixels.
By definition that would be clusters, but from several particles and not from one.
To evaluate the average cluster size from individual particles, the Gaussian distribution of the cluster sizes is fitted and the baseline extracted.
This results in a baseline average cluster size of 1.14 pixels for carbon ions of a wide energy range from \SIrange{108}{423}{\mega\electronvolt\per u}.
For protons with energies in the range of \SIrange{60}{217}{\mega\electronvolt} the average cluster size is 1.12 pixels.\\
These results confirm the correct selection of the pixel dimensions.

\subsection{Radiation Tolerance}
\label{sec:radiation_tolerance}
\begin{figure}
    \centering
    \includegraphics[width=0.8\textwidth]{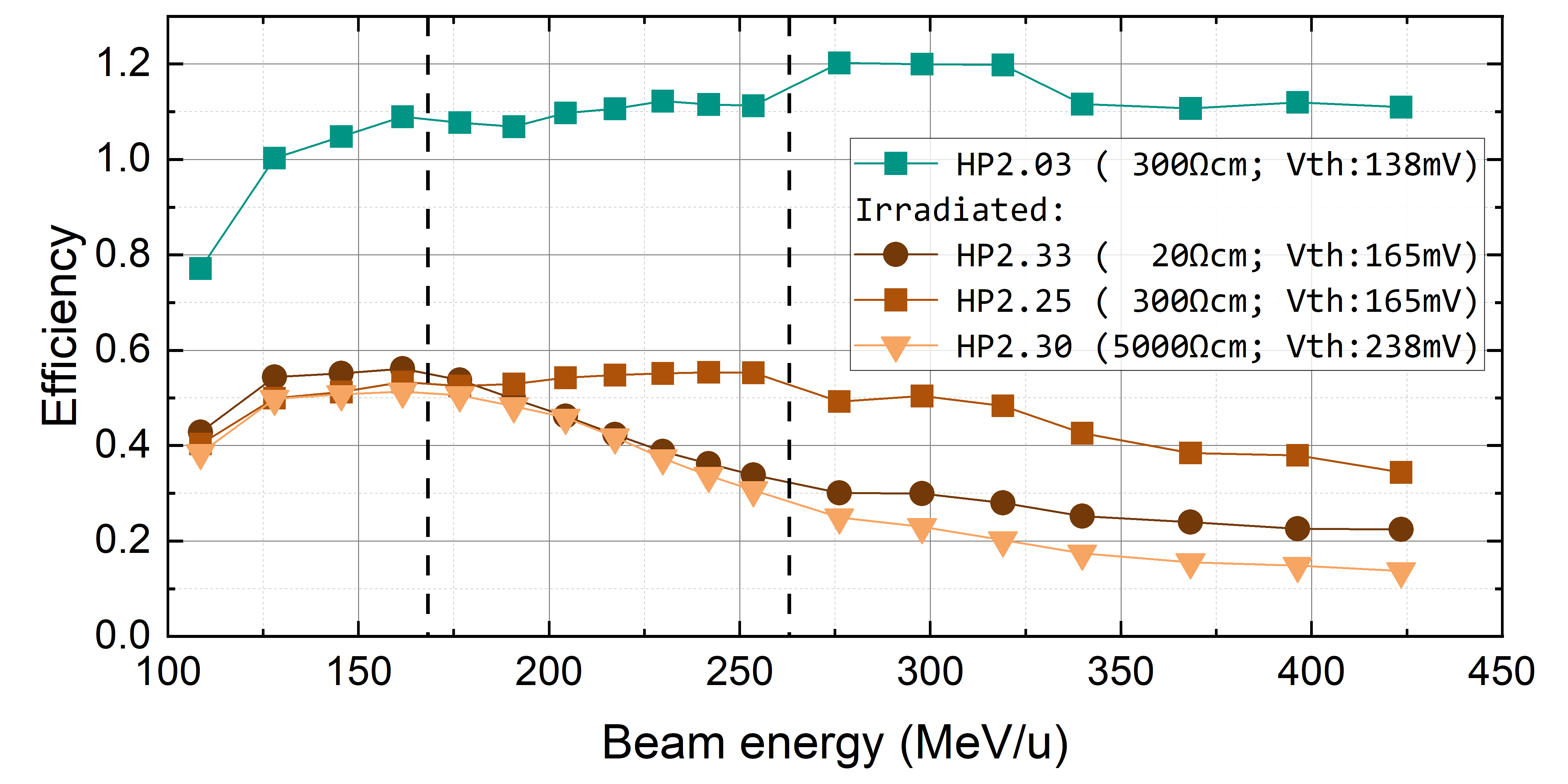}
    \caption{Hit detection efficiency measured with carbon ions of different energies after~\cite{pittermann_EvaluationHVCMOSSensors_2022}. Sample \texttt{HP2.03} (\SI{300}{\ohm\centi\meter}) is not irradiated and operated at \SI{50}{\volt}, while the other samples are irradiated to a \SI{1}{\mega\electronvolt} equivalent fluence of \SI{1e15}{\per\square\centi\metre} and operated at \SI{25}{\volt}. 
    The two numbers in the brackets behind the sample name in the legend correspond to substrate resistivity and comparator threshold voltage (Vth), respectively.
    Beam parameters: carbon ions, \SI{5e6}{\per\second}/\SI{3e6}{\per\second} particle rate for the unirradiated/irradiated sample, respectively, spot width adapted (steps indicated by vertical dashed lines) depending on the beam energy to stay at around \SIrange{8}{9}{\milli\meter} (FWHM).}
    \label{fig:carbon_homog}
\end{figure}

Detectors with different substrate resistivities were irradiated at ZAG with \SI{23}{\mega\electronvolt} protons up to a \SI{1}{\mega\electronvolt} neutron equivalent fluence of \SI{1e15}{\per\square\centi\metre}.
The uniform irradiation resulted in an increase of the leakage current as listed in Table~\ref{tab:currents_irrad}.
\begin{table}[]
    \caption{Tentative currents for the \SI{1}{\square\centi\metre} detectors of different substrate resistivities. The current measurements were performed at room temperature and at \SI{25}{\volt} bias voltage.}
    \centering
    \begin{tabular}{c|c|c}
        \toprule
        Substrate resistivity & Current before irrad. & Current after irrad. \\
        \midrule
        \SI{20}{\ohm\centi\meter} & <\SI{100}{\nano\ampere} & \SI{220}{\micro\ampere}\\
        \SI{300}{\ohm\centi\meter} & <\SI{100}{\nano\ampere} & \SI{130}{\micro\ampere}\\
        \SI{5000}{\ohm\centi\meter} & <\SI{100}{\nano\ampere} & \SI{80}{\micro\ampere}\\        
        \bottomrule
    \end{tabular}
    \label{tab:currents_irrad}
\end{table}
These irradiated detectors were exposed to a carbon ion beam at HIT with a defined intensity from which the total number of particles is derived (uncertainty of 5\% on number of particles in the QA room; less than 2\% in the treatment rooms).
This number is compared to the number of acquired hits and the ratio is the hit detection efficiency quoted in Figure~\ref{fig:carbon_homog}.
The hit detection efficiency of the unirradiated sample \texttt{HP2.03} exceeds 100\% since for each particle on average 1.14 hits are registered (Section~\ref{sec:hit_detection}).
The uniformly irradiated detectors still work and detect the beam spot correctly. 
Compared to the unirradiated reference sample, the efficiency is reduced, which gets more pronounced at higher ion energies, for which less ionization and, therefore, signal is expected.
Since the results are based only on one sample each, one cannot draw a final conclusion on the best choice of substrate resistivity, yet.
There are further test conditions which have not been optimized so far.
The irradiated samples were operated at room temperature and not cooled which can lower the leakage current and noise.
Therefore, also the threshold was possibly set too high so that the pixels were less sensitive to small signals.
A further handle to improve the efficiency is a higher bias voltage, which was limited to \SI{25}{\volt} due to the high leakage current of the samples with the lowest resistivity in this comparative study.\\ 
In general, the detectors are operational after irradiation up to a fluence equivalent to one year operation at the central spot, but further tuning is required to keep the efficiency high avoiding the need for too frequent calibrations.

\subsection{Beam Monitoring}
\begin{figure}
    \centering
    \includegraphics[width=0.5\textwidth]{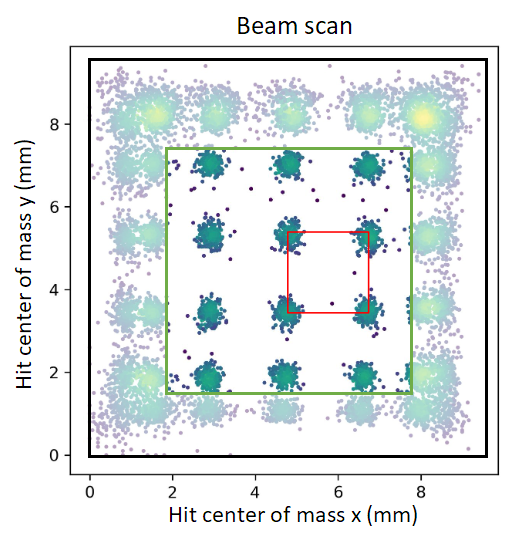}
    \caption{Reconstructed beam centers for a continuous spot scan with \SI{2}{\milli\meter} steps. The detector was in counter readout mode with \SI{50}{\micro\second} integration time per frame. Each point in the plot corresponds to the reconstructed beam center in a frame. The \SI{2}{\milli\meter} raster is indicated by the red square. The green square indicates the area without distortions which occur at the edge of a single detector. Beam parameters: carbon ions, \SI{430}{\mega\electronvolt\per u} energy, \SI{3.4}{\milli\meter} spot width (FWHM), \SI{5e6}{\per\second} particle rate.}
    \label{fig:BeamScan2mm}
\end{figure}

\begin{figure}
    \centering
    \includegraphics[width=0.75\textwidth]{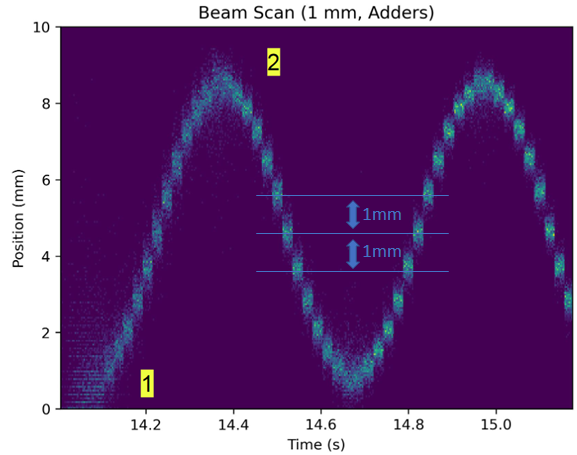}
    \caption{Reconstructed beam centers for a continuous spot scan with \SI{1}{\milli\meter} step size using adder readout with \SI{20}{\micro\second} integration time. Two features are indicated in the plot: at '1' the beam intensity ramps up, and at '2' part of the beam spot extends beyond the detector edge, which deteriorates the position reconstruction. In the center the beam spot is contained within the detector matrix and a correct reconstruction is possible showing the expected \SI{1}{mm} steps. Beam parameters: carbon ions, \SI{430}{\mega\electronvolt\per u} energy, \SI{3.4}{\milli\meter} spot width (FWHM), \SI{5e6}{\per\second} particle rate.}
    \label{fig:BeamScan1mmAdder}
\end{figure}

\begin{figure}
    \centering
    \includegraphics[width=0.85\textwidth]{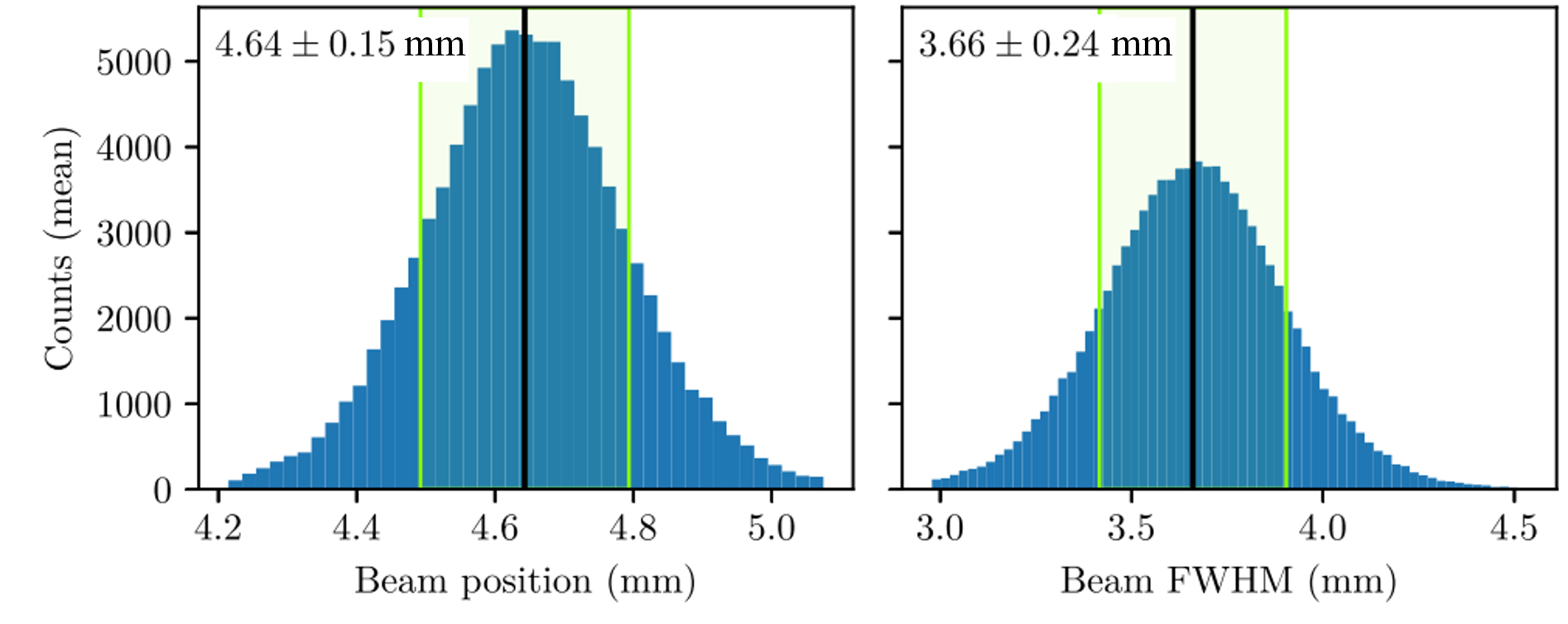}
    \caption{Reconstructed beam centers (left) and beam widths (right) for a steady carbon ion beam for projected hit profiles from~\cite{pittermann_EvaluationHVCMOSSensors_2022}. The integration time for a frame used for reconstruction was \SI{30}{\micro\second}. Beam parameters: carbon ions, \SI{430}{\mega\electronvolt\per u} energy, \SI{5e6}{\per\second} particle rate, \SI{3.4}{\milli\meter} spot width (FWHM).}
    \label{fig:steady_beam_hist}
\end{figure}

Fast reconstruction of the beam position is one of the key requirements for a beam monitoring system.
With the current HitPix2 detector only an area of $\SI{9.6}{\milli\metre}\times\SI{9.6}{\milli\metre}$ is sensitive for particles and, therefore, only very tiny beam spots and small movements can be traced.
Distortions of the reconstruction at the edges of the detector are expected and will not occur in the final system, since the missing beam spot part will be recorded by the surrounding detectors.
Figure~\ref{fig:BeamScan2mm} shows the reconstructed beam positions for a carbon ion beam with the smallest focus value possible at HIT.
Between each irradiation spot, the beam is moved by \SI{2}{\milli\metre}.
During one spot irradiation of several seconds length, the position is reconstructed many times from the hits acquired during a \SI{50}{\micro\second} frame, and each of the reconstructed points is one entry in the histogram of Figure~\ref{fig:BeamScan2mm}.
For the reconstruction of the beam spot, a simple mean algorithm is used, which can also be easily implemented on an FPGA.
The projected hits are binned by pixels at positions $x_i$ with number of hits $n_i$.
The mean beam position $x_0$ and beam width $\sigma_0$ (or full width half maximum FWHM) are calculated as
\begin{linenomath}
\begin{align}
    x_0 &= 
        \frac{1}{n_\mathrm{tot}}
        \sum_{i} n_i \cdot x_i \label{eq:reco_mean_pos} \\
    \sigma_0^2 &=
        \frac{1}{n_\mathrm{tot} - 1}
        \sum_{i} n_i \cdot \left(x_i - x_0\right)^2 \quad (\mathrm{FWHM} = 2 \sqrt{2 \ln{2}}\;  \sigma_0). \label{eq:reco_mean_width}
\end{align}
\end{linenomath}
In the center, one can clearly see the separation of \SI{2}{\milli\metre} as intended for this irradiation plan.
At the detector edges the reconstructed position is distorted due to the missing particles outside the detector as expected.\\
A special feature of the HitPix family is the projection of the hit counts on one (later two) axis (adder readout mode).
Figure~\ref{fig:BeamScan1mmAdder} illustrates this feature.
It shows the reconstructed positions of the projected beam profile integrated within \SI{20}{\micro\second} during a spot scanning irradiation. 
The small carbon ion beam spot was scanned in \SI{1}{\milli\metre} steps.
This demonstrates that in this acquisition mode, which will be the default during operation as beam monitor, the beam motion can be traced continuously without dead time during readout.\\
To investigate the precision of the reconstruction method, a steady beam of carbon ions was monitored for several seconds and the reconstructed position and width filled in the two histograms in Figure~\ref{fig:steady_beam_hist}.
The standard deviation or precision of the reconstructed position in this case is \SI{150}{\micro\meter}.
Assuming a chip alignment precision of \SI{10}{\um}, which is a plausible value assuming an automatized pick and place procedure, the resulting position accuracy is below the required accuracy of \SI{200}{\micro\meter}.
For the presented low intensity of \SI{5e6}{\per\second} and \SI{30}{\micro\second} integration time, each projected beam profile contains about 150 hits on average, which limits the reconstruction precision.
Higher intensities result in a better precision, e.g. at \SI{5e7}{\per\second} the precision is improved to \SI{80}{\micro\meter}.\\
The width of the beam is taken as the standard deviation of the hit distribution.
From these measurements, a mean beam width of \SI{3.66}{\milli\meter} (FWHM) is derived and indicated in the Figure~\ref{fig:steady_beam_hist} (right).
This is slightly higher than the nominal width of \SI{3.4}{\milli\meter}, but still well within the allowed delivered width range of +25\%/--10\%.\\
In this section, we have demonstrated that with the HitPix detector it is possible to monitor the beam position and width with the required accuracy.
The short acquisition time of below \SI{30}{\micro\second} allows rapid feedback, and the operation in projection mode allows fast readout of the data, and, thus, low latency of the results.

\subsection{High-Rate Capability}
\begin{figure}
    \centering
    \includegraphics[width=0.85\textwidth]{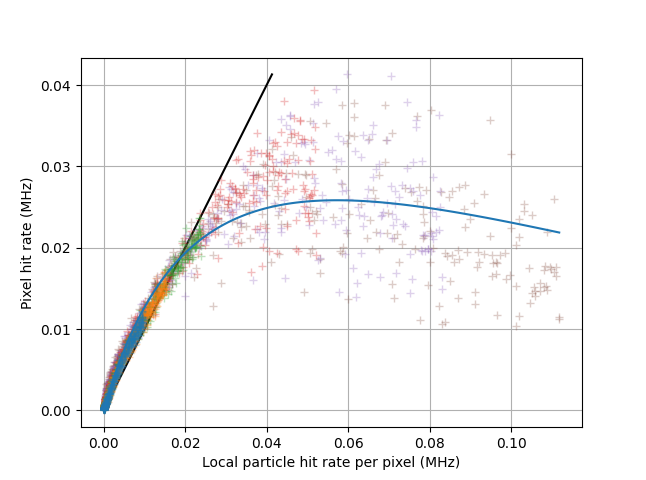}
    \caption{Single pixel hit rates versus expected local particle rates from ~\cite{pittermann_EvaluationHVCMOSSensors_2022}.  The plot combines data from spot irradiations with different beam intensities (colors: \SI{5e6}{\per\second} (blue), \SI{1e7}{\per\second} (orange), \SI{2e7}{\per\second} (green), \SI{3e7}{\per\second} (red), \SI{5e7}{\per\second} (violet), \SI{8e7}{\per\second} (brown)). For each pixel, the expected local particle rate is calculated assuming a Gaussian beam profile and the nominal beam intensities. The black line indicates a 100\% hit detection efficiency. The blue line is a fit according to Equation~\ref{eq:Takacs} with $\epsilon=1.75$, $\theta=0.31$ and $\tau=$\SI{20.6}{\micro\second}.  Beam parameters: carbon ions, \SI{430}{\mega\electronvolt\per u} energy, \SI{3.4}{\milli\meter} spot width (FWHM).}
    \label{fig:scatter_colors}
\end{figure}

\begin{figure}
    \centering
    \begin{subfigure}[b]{0.58\textwidth}
        \includegraphics[width=\textwidth]{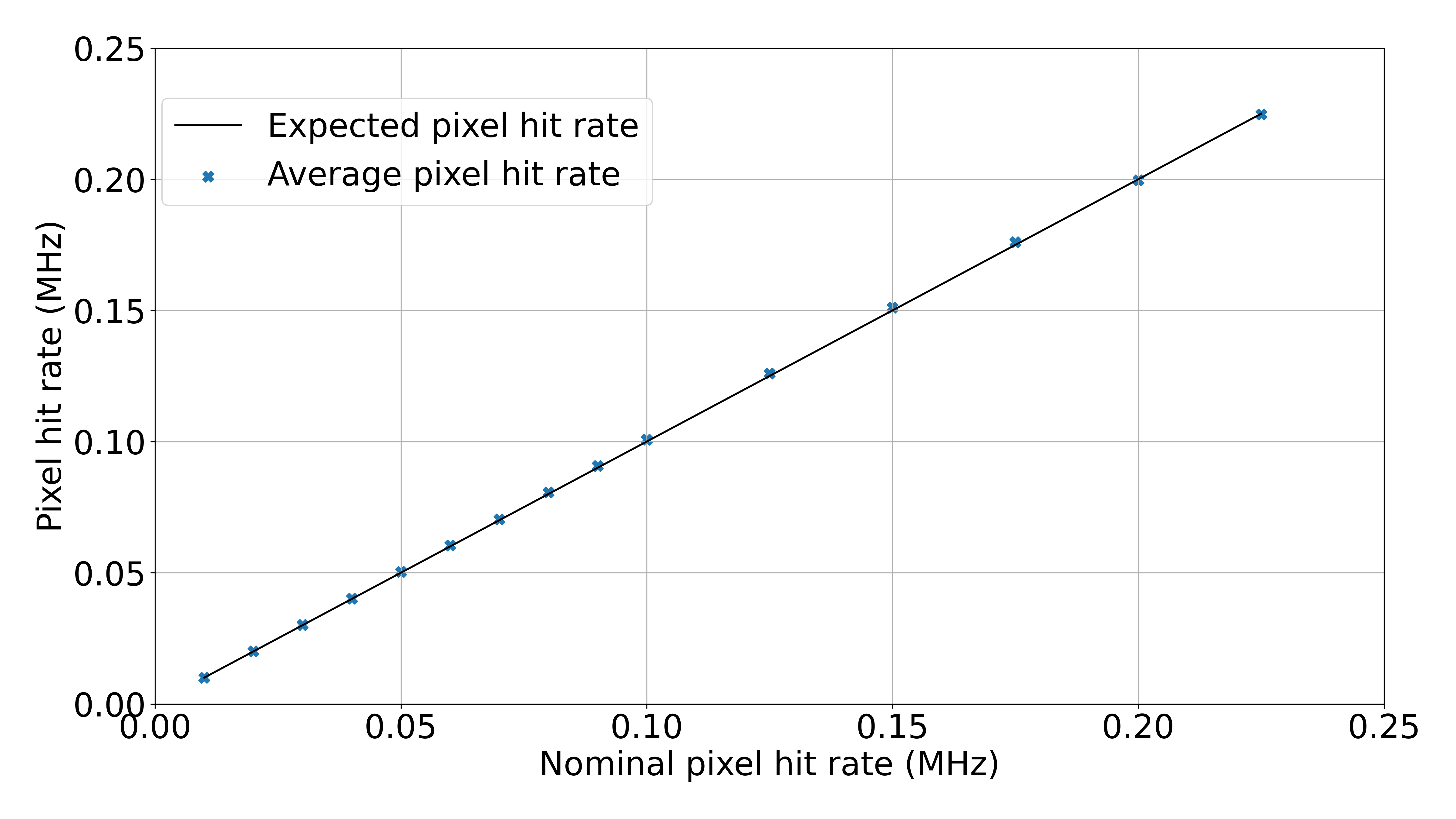}
        \caption{Average hit rate for the pixels in the red rectangle on the right}
    \end{subfigure}\hfill
    \begin{subfigure}[b]{0.38\textwidth}
        \includegraphics[width=\textwidth]{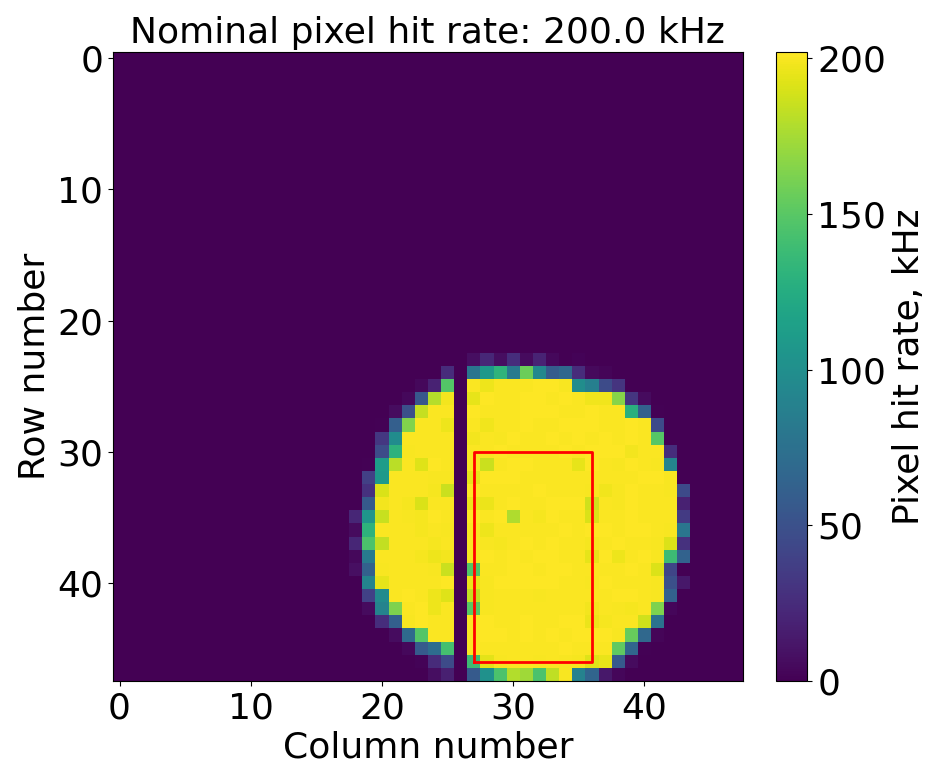}
        \caption{Map of the pixel hit rates for \SI{200}{\kHz} injections}
    \end{subfigure}
    \caption{Hit rates of pixels illuminated by a pulsed infrared LED. The average hit rate (blue symbols) versus the expected nominal hit rate is plotted in (a). The black line indicates the expected rate for 100\% efficiency. (b) shows, as an example, the map of the pixel rates for \SI{200}{\kHz} injections.}
    \label{fig:led}
\end{figure}

\begin{figure}
    \centering
    \begin{subfigure}[b]{0.32\textwidth}
        \includegraphics[width=\textwidth]{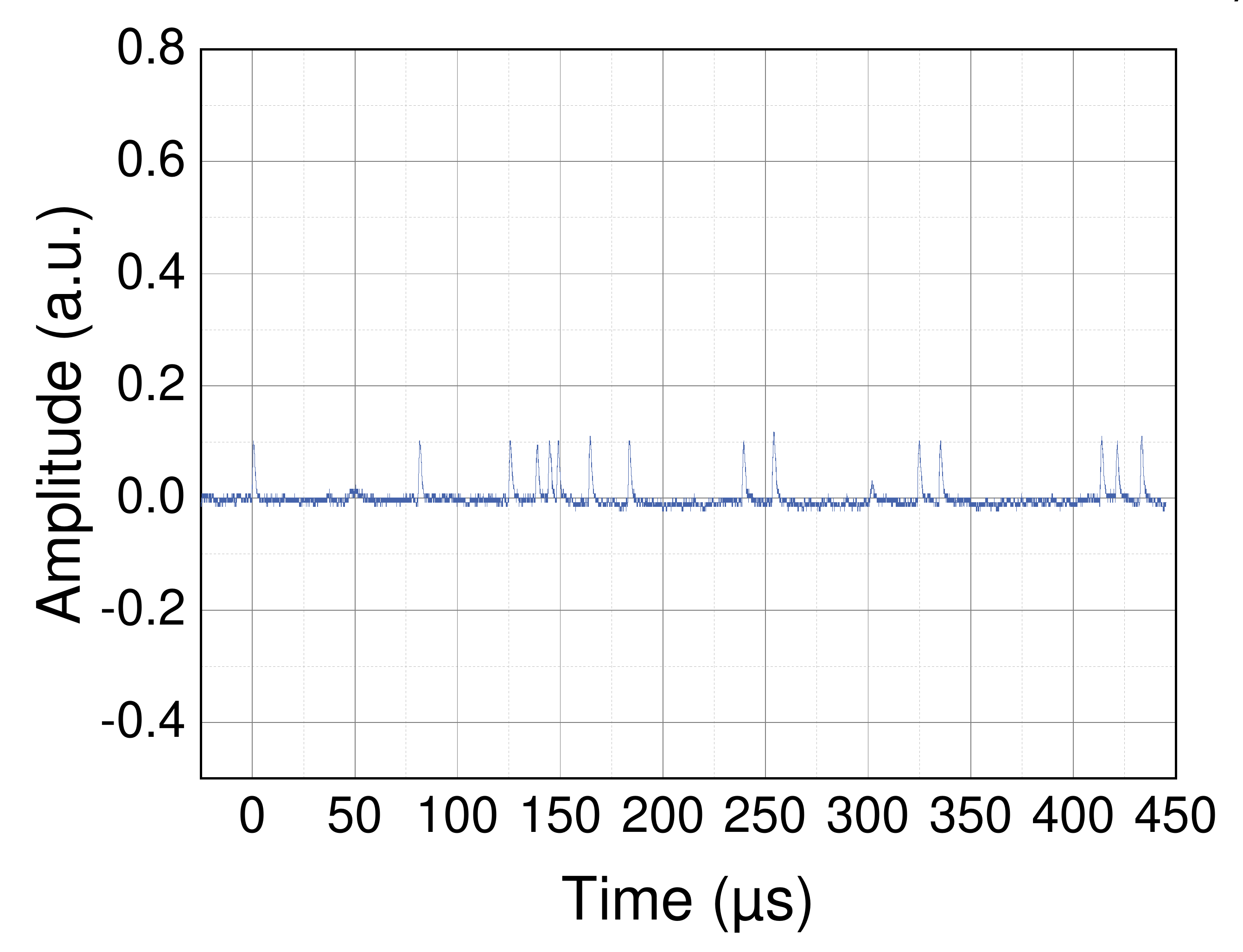}
        \caption{Signals for protons at a rate of \SI{8e7}{\per\second}.}
    \end{subfigure}\hfill
    \begin{subfigure}[b]{0.32\textwidth}
        \includegraphics[width=\textwidth]{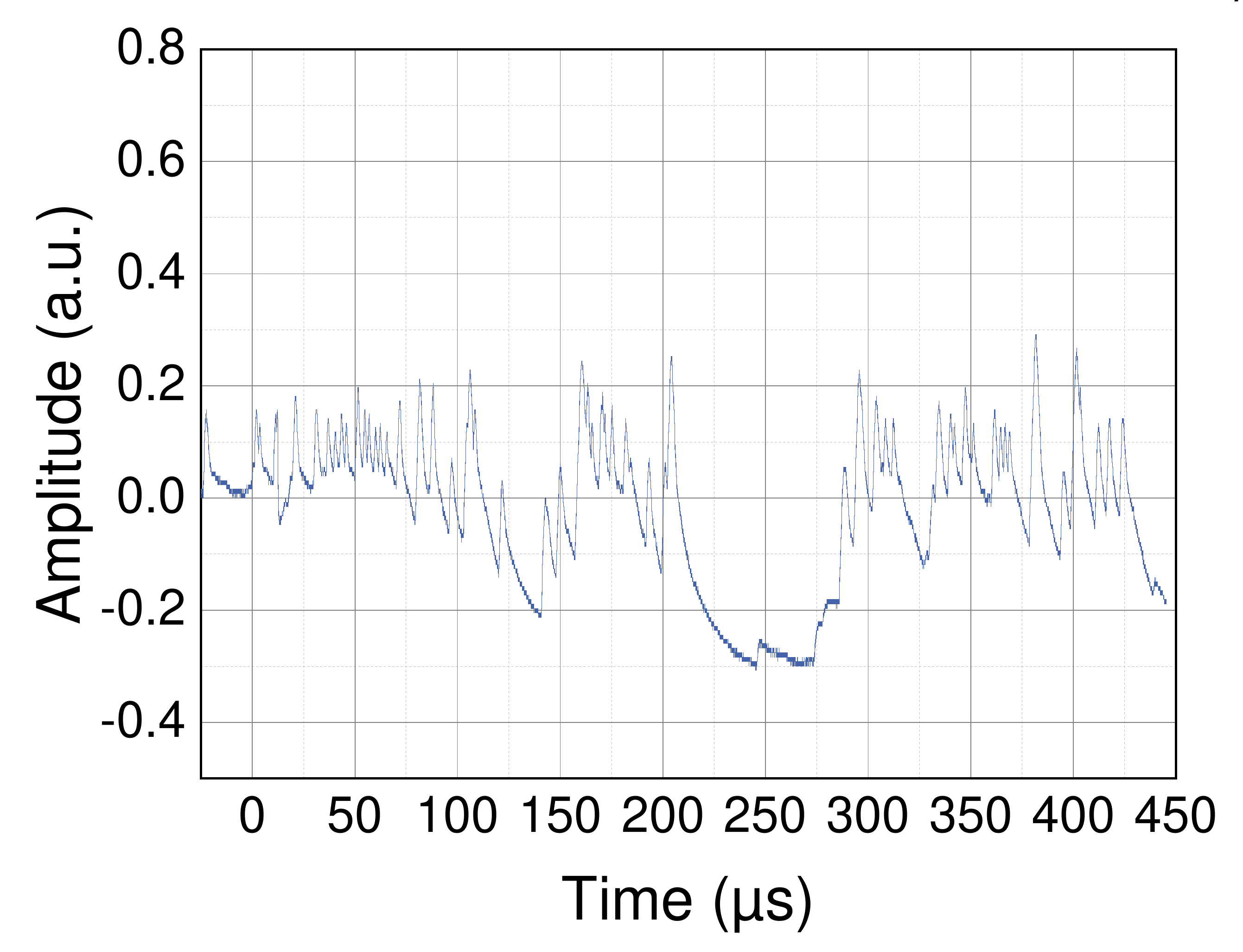}
        \caption{Signals for protons at a rate of \SI{4e8}{\per\second}.}
    \end{subfigure}\hfill%\vspace{5mm}
    \begin{subfigure}[b]{0.32\textwidth}
        \includegraphics[width=\textwidth]{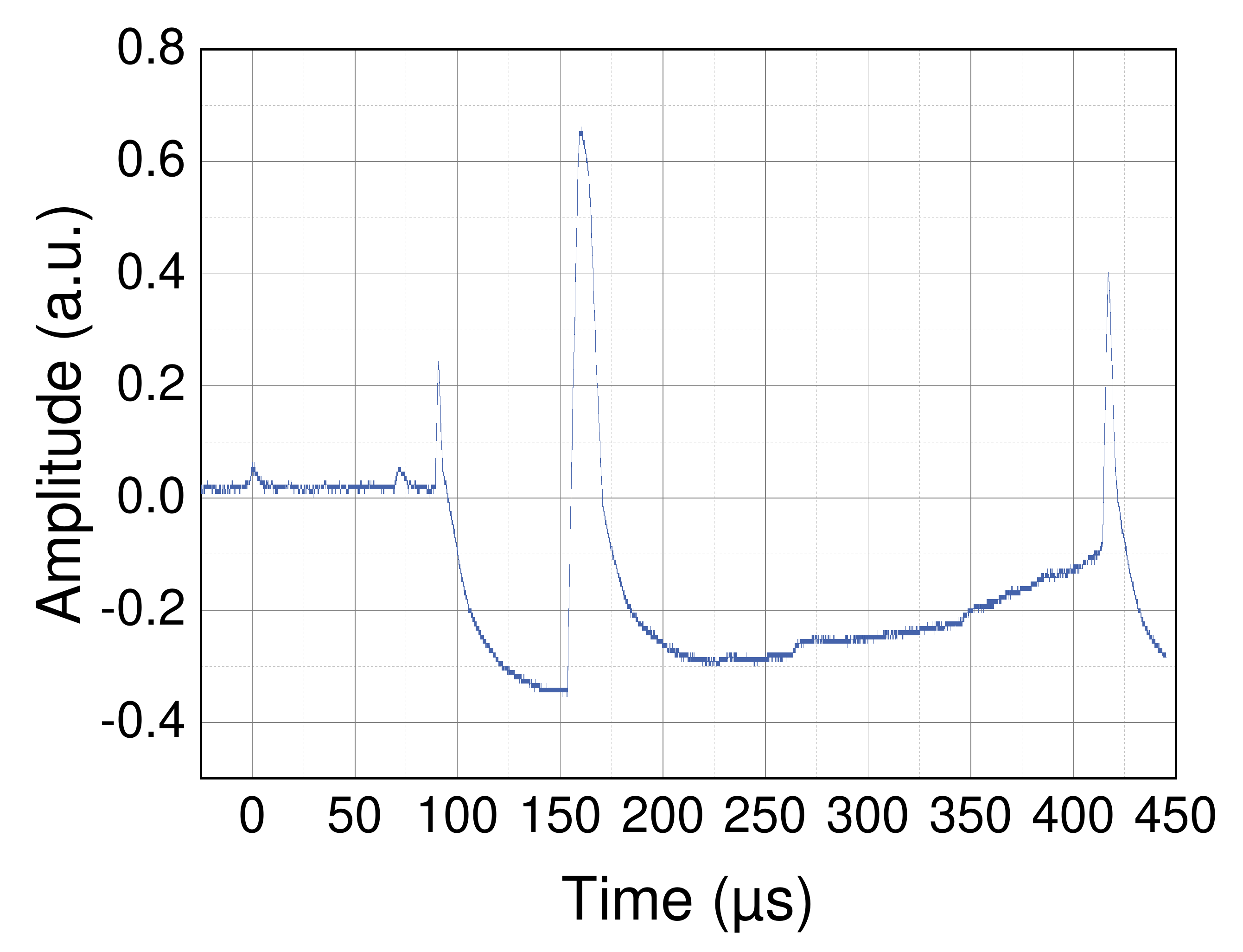}
        \caption{Signals for carbon ions at a rate of \SI{2e6}{\per\second}.}
    \end{subfigure}
    \caption{Examples of voltage measurements at the amplifier output of one pixel when exposed to different particle beams.}
    \label{fig:AmpOut}
\end{figure}

The measurements presented so far have been mainly acquired at low beam intensities.
This section focuses on the high-rate capability of the HitPix2 detector.
To address this property, an intensity scan of a steady and small carbon ion beam at highest energy was investigated.
The expected local particle rate is calculated by the average intensity of the spot irradiation and the position of the pixel with respect to the beam center assuming a Gaussian beam profile.
The registered hit rate versus the expected local particle rate for each pixel at several beam intensities is plotted in Figure~\ref{fig:scatter_colors}.
At low intensities, the pixels register the expected number of hits also considering that, on average, each particle generates 1.14 hits (Section~\ref{sec:hit_detection}).
With increasing local particle rate the registered hit rate saturates and a huge spread for individual pixels is observed.
This effect can be modeled taking dead time of the hit detection into account using a slightly modified version of Takacs' formula~\cite{muller_GeneralizedDeadTimes_1991}:
\begin{linenomath}
\begin{equation}\label{eq:Takacs}
    \nu_\mathrm{out} = \dfrac{\epsilon \cdot \nu \cdot \theta}{e^{\epsilon \cdot \nu \cdot \theta \cdot \tau}+\theta -1}    
\end{equation}
\end{linenomath}
where $\nu_\mathrm{out}$ is the measured hit rate, $\nu$ the true particle rate, $\epsilon$ the hit efficiency (additional parameter to account for clusters), $\theta$ the probability of paralyzing dead time ($\theta=1$ would be fully paralyzing, i.e. hits during dead time extent the dead time), and $\tau$ the dead time.
The blue line in Figure~\ref{fig:scatter_colors} represents the result of a fit to the data indicating a dead time of $\tau=$~\SI{20.6}{\micro\second}.
This behavior is unexpected both from design and laboratory measurements and is not in compliance with the requirements.
If the registered hit rate would follow a steady function with respect to the local particle rate, the hit rate could be corrected, but for dropping and widely scattered rates this is not possible.\\
Further investigations have been performed to understand this behavior.
Short light pulses generated by a pulsed infrared LED were injected on the detector with increasing pulse frequency.
The pixel counts were compared to the number of light pulses.
For optimally illuminated pixels the count rate follows the injection rate up to a rate of \SI{225}{\kHz} as shown in Figure~\ref{fig:led}.
In this configuration the hit detection efficiency is very high up to much higher injection rates than in the ion beam in Figure~\ref{fig:led}.
This indicates that the fast counting capability of the pixels is functioning.\\
Looking at the analog signals of the amplifier of a pixel in the ion beam we observe an increasing number of dips of the baseline voltage when exposed to higher particle rates.
Figure~\ref{fig:AmpOut} shows examples of such measurements.
Carbon ions generate the most severe, but less frequent dips.
During the time the baseline is reduced, no hits are detected, causing dead time.\\
These measurements suggest that the heavily ionizing ions influence the functioning of the pixels when a certain region in the pixel cell is hit.
In addition, analog signals were recorded of injected alpha particles with an energy of about \SI{5.5}{\MeV} from an Am-241 source.
Similar baseline voltage drops as in the ion beam could be provoked with these alpha particles.
By placing the source above different locations of the detector, it could be excluded that the electronics in the periphery is affected.\\
In conclusion, the hit detection efficiency of the HitPix2 drops for high-rate particle beams.
The investigations indicate that at a high rate of highly ionizing particles the baseline voltage is affected and hit detection is deteriorated.
Still, the pixel cells function as expected for moderate charge injection and further work can concentrate on mitigating effects induced by highly ionizing particles.
If the high-rate capability is not improved in a following version of the detector the beam monitoring requirements cannot be met.

\subsection{Magnetic Field Tolerance}\label{sec:Studies_MRI}
Magnetic fields are required for MRI-guided radiotherapy, however, they influence the movement of charged particles and charge carriers in solids.
Since all modern detector technologies in beam monitoring rely at least partially on electric fields and movement of charge carriers, the operation within the strong magnetic field of an MRI machine or the somewhat lower fringe field of a scanner has to be part of the technology evaluation process.\\
In a first step, we have investigated the performance of a detector setup with HitPix1 in the adjustable field of the Helmholtz coils as described in Section~\ref{sec:Systems_MRI}.
Unfortunately, but as expected, the commercial FPGA board has several components (mainly voltage regulation) which cannot be operated in magnetic fields.
Therefore, the readout stopped already at \SI{60}{\milli\tesla} field. 
Up to this field strength, the detector kept working and no performance degradation in monitoring a carbon ion beam has been detected.\\
To cope with the sensitivity of the FPGA board for these tests, a PCIe x16 twisted-pair riser cable with 1:1 connectivity has been designed and produced.
Using this cable, the commercial readout electronics was moved out of the field center and the detector readout worked up to \SI{100}{\milli\tesla} which is the maximum possible field of the Helmholtz-coil pair.\\
For MRI tests, the modified setup with riser cable and HitPix2 has been used.
The detector carrier-PCB has been placed in the middle of the MRI (described in Section~\ref{sec:Systems_MRI}) and the readout FPGA board at a distance of approximately \SI{1}{\meter}.
No influence of the static magnetic field was observed in this configuration.
The rapidly changing field of the scanning MRI, however, causes the detector to lose its configuration, rendering it non-operational. 
The reason for the configuration loss is the large induced voltage on the detector's configuration lines.
Depending on the MRI scanning procedure, we measured pulses of several volts on these lines, while the logic levels are about \SI{1.8}{\volt}.
At a distance of \SI{35}{\centi\meter} from the center of the scanning MRI (approximately one quarter of the central field strength), the detector kept its configuration and readout worked normal.\\
In the next iteration of the detector, a lockable configuration is foreseen, meaning that random noise on the configuration line will be ignored and only configuration bits preceded by an unlock-code will be accepted.

\subsection{Multi-Detector Readout}
\begin{figure}
    \centering
    \includegraphics[width=0.6\textwidth]{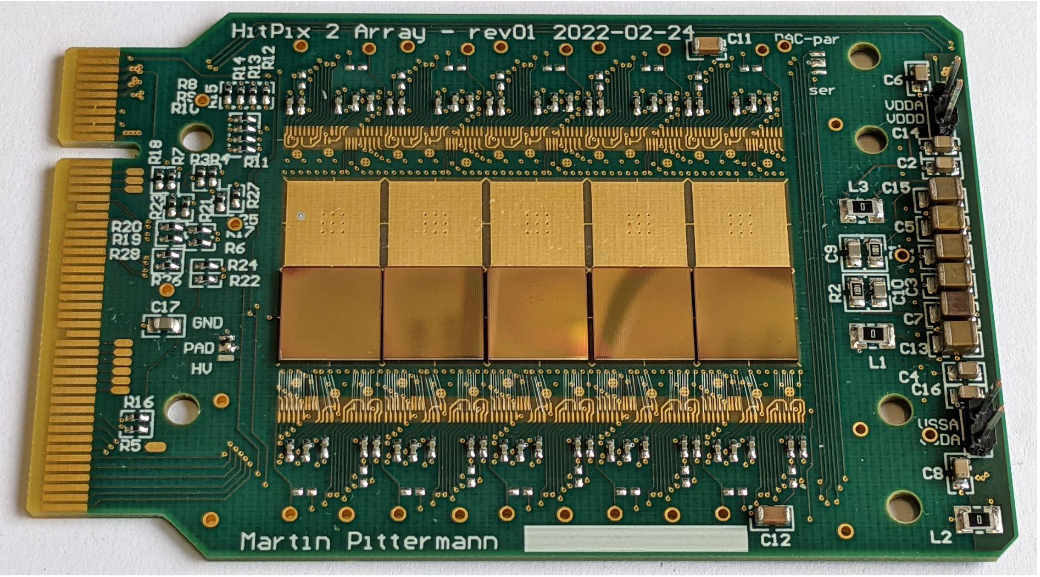}
    \caption{Picture of the $2 \times 5$ matrix PCB with five mounted detectors.}
    \label{fig:MatrixFoto}
\end{figure}

\begin{figure}
    \centering
    \begin{subfigure}[b]{0.9\textwidth}
        \includegraphics[width=0.95\textwidth]{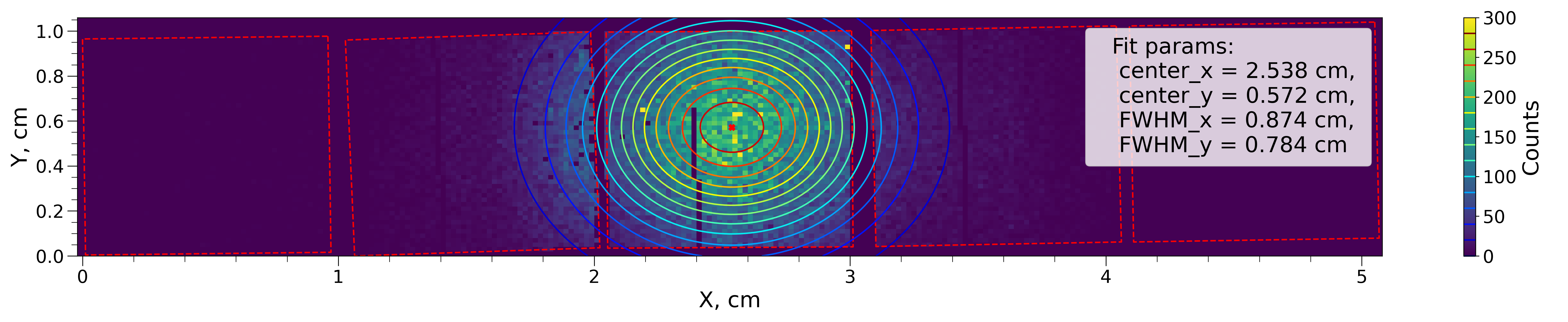}
        \caption{Protons with beam width of \SI{0.81}{\centi\meter} (FWHM)}
    \end{subfigure}\vspace{5mm}
    \begin{subfigure}[b]{0.9\textwidth}
        \includegraphics[width=0.95\textwidth]{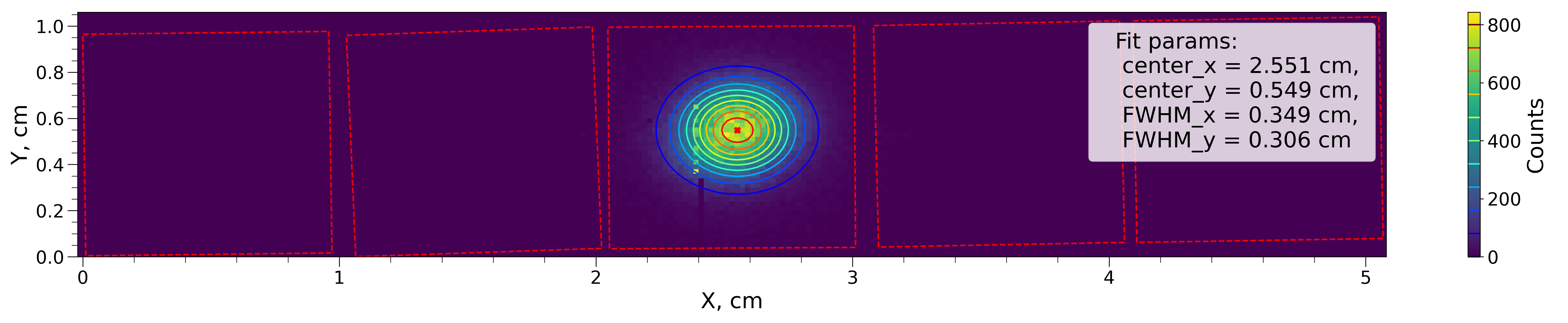}
        \caption{Carbon ions with beam width of \SI{0.34}{\centi\meter} (FWHM)}
    \end{subfigure}\vspace{5mm}
    \begin{subfigure}[b]{0.9\textwidth}
        \includegraphics[width=0.95\textwidth]{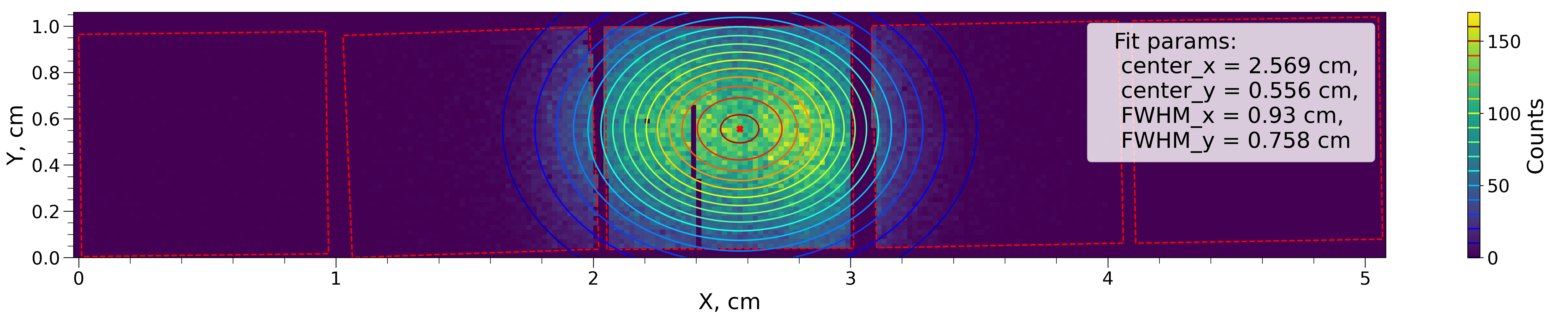}
        \caption{Carbon ions with beam width of \SI{0.78}{\centi\meter} (FWHM)}
    \end{subfigure}
    \caption{Hit maps for all five detectors of the matrix after alignment for three different beam settings. 
    The red squares indicate the sensitive area of the aligned detectors. 
    The colored circles are contour lines of the 2d-Gaussian profile fitted to the hit distribution.
    The fit parameters are provided in the plots.
    Each plot contains about \num{250000} hits.}
    \label{fig:MatrixHitmap}
\end{figure}

To reach the required large size of the beam monitor, it is necessary to read out many detectors.
HitPix2 implements two schemes for multi-detector readout. 
In daisy chain mode, several detectors can be directly interconnected (digital \texttt{data\_out} to \texttt{data\_in}) and the data shifted through all of them.
The other mode is a bus configuration for which several detectors can be connected to a common data line and the detectors individually selected for readout.\\
A PCB was designed to validate these functionalities in a $2 \times 5$ matrix (Figure~\ref{fig:MatrixFoto}).
A first PCB was equipped with five detectors in one row.
The detectors were successfully read out in both readout modes.
At HIT, the matrix was exposed to particle beams for different beam parameter settings.
The collected hit counts for three beam settings are shown in Figure~\ref{fig:MatrixHitmap}.
The reconstructed beam spot widths are well in agreement with the nominal settings. 
Especially in the bottom plot one can see a smooth transition of hit counts from one detector chip to the next.
This demonstrates the comparable hit detection efficiency of the neighboring detectors.\\
To achieve very high readout rates the bus configuration should be used since the daisy chain readout showed a limit of about \SI{28}{\mega bit \per \second} on the test setup.

\subsection{Simulation Results}
\begin{figure}
    \centering
    \includegraphics[width=.8\textwidth]{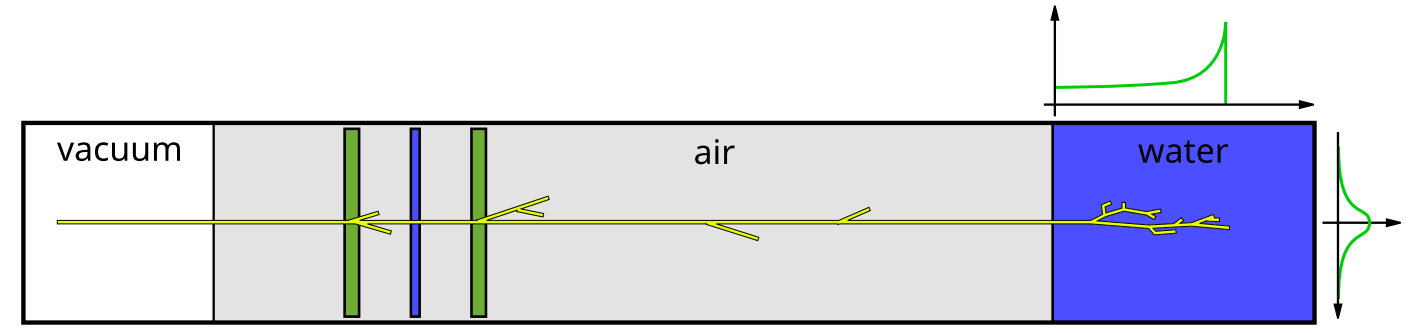}
    \caption{Illustration of the simulated setup. The green rectangles represent the position sensitive detectors and the narrow blue rectangle between them the effective water equivalent of the ionization chambers. The plots at the water box indicate examples of the expected transverse (top plot) and lateral (right plot) energy deposition.}
    \label{fig:sim_scene}
\end{figure}

\begin{figure}
    \centering
    \includegraphics[width=.9\textwidth]{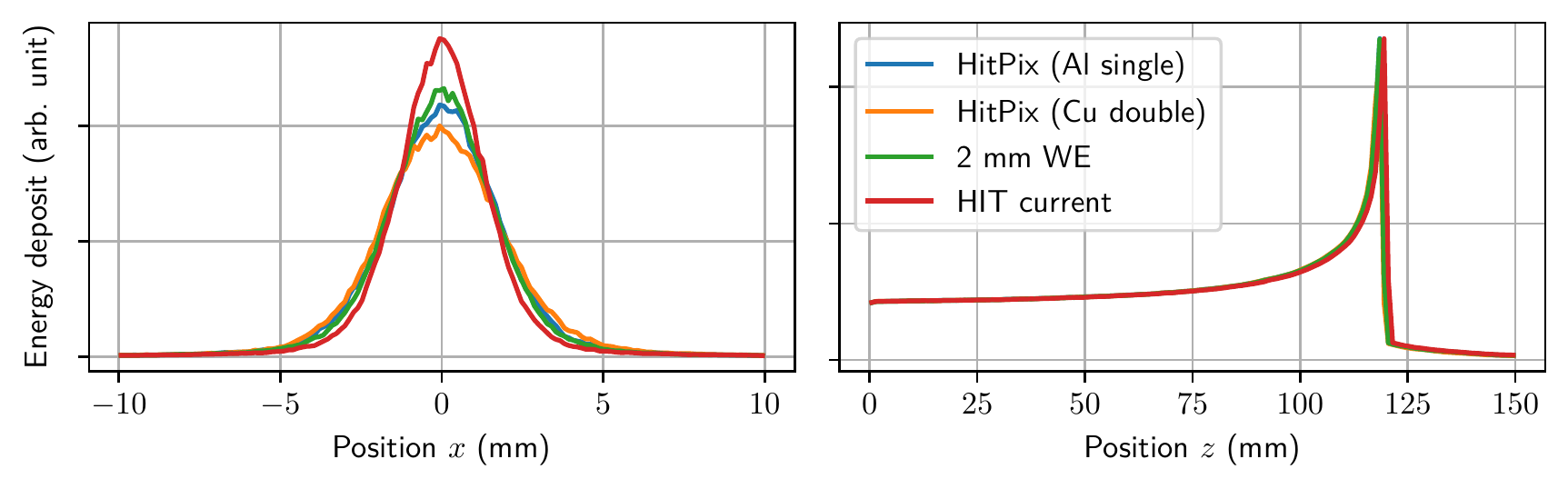}
    \caption{Comparison of the simulation results for the different arrangements from~\cite{pittermann_EvaluationHVCMOSSensors_2022}. Plotted are the lateral (left) and transverse (right) energy deposition.}
    \label{fig:scenes_compare}
\end{figure}

\begin{figure}
    \centering
    \includegraphics[width=.8\textwidth]{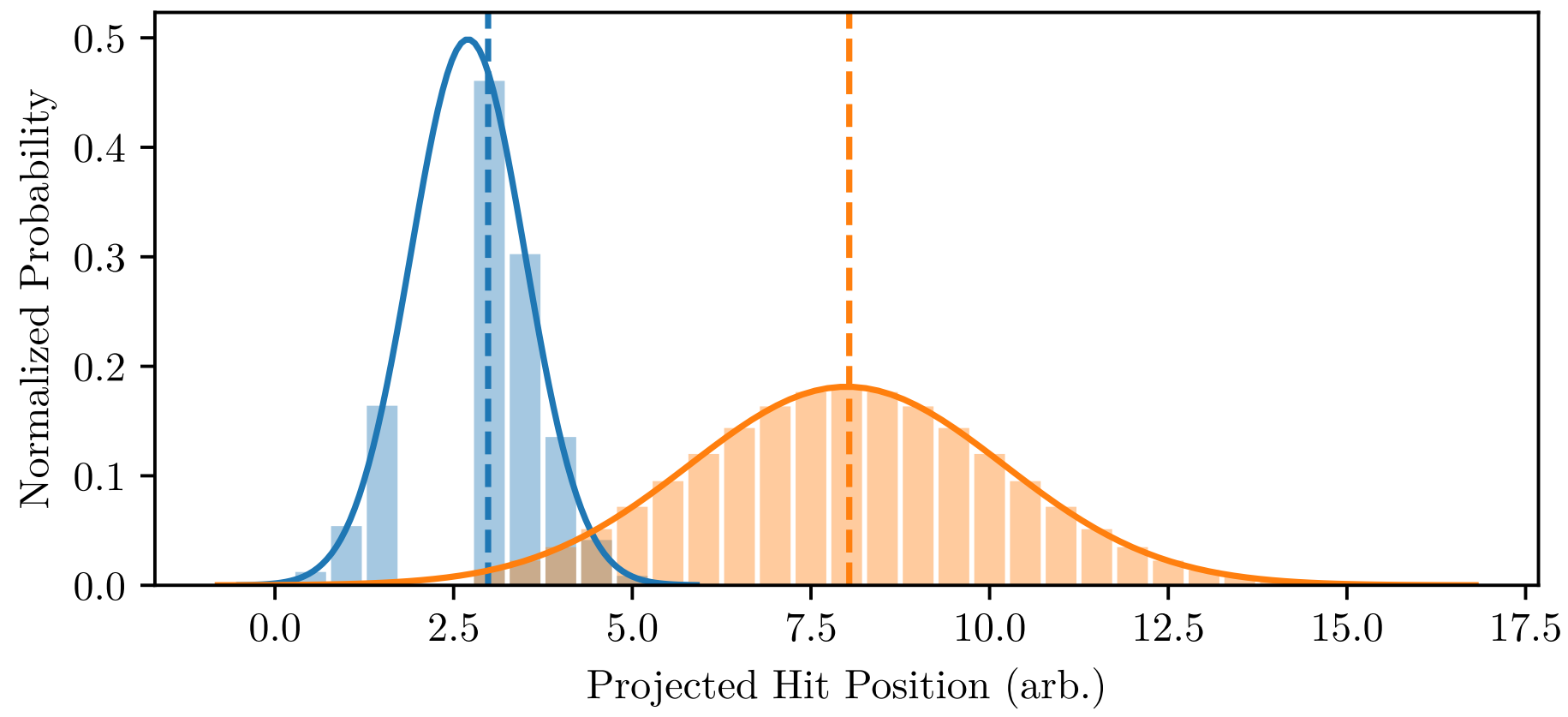}
    \caption{Illustration of two simulated beam profiles with an insensitive gap at $x=2.5$ from~\cite{pittermann_EvaluationHVCMOSSensors_2022}. The vertical dashed lines represent the true center of the beams, and the solid lines a Gaussian fit of the hit counts represented by the bars. For the narrow beam, the bias of the reconstructed peak is visible. Statistical fluctuations are omitted for this example.}
    \label{fig:sim_gap_illustration}
\end{figure}

\begin{figure}
    \centering
    \includegraphics[width=.8\textwidth]{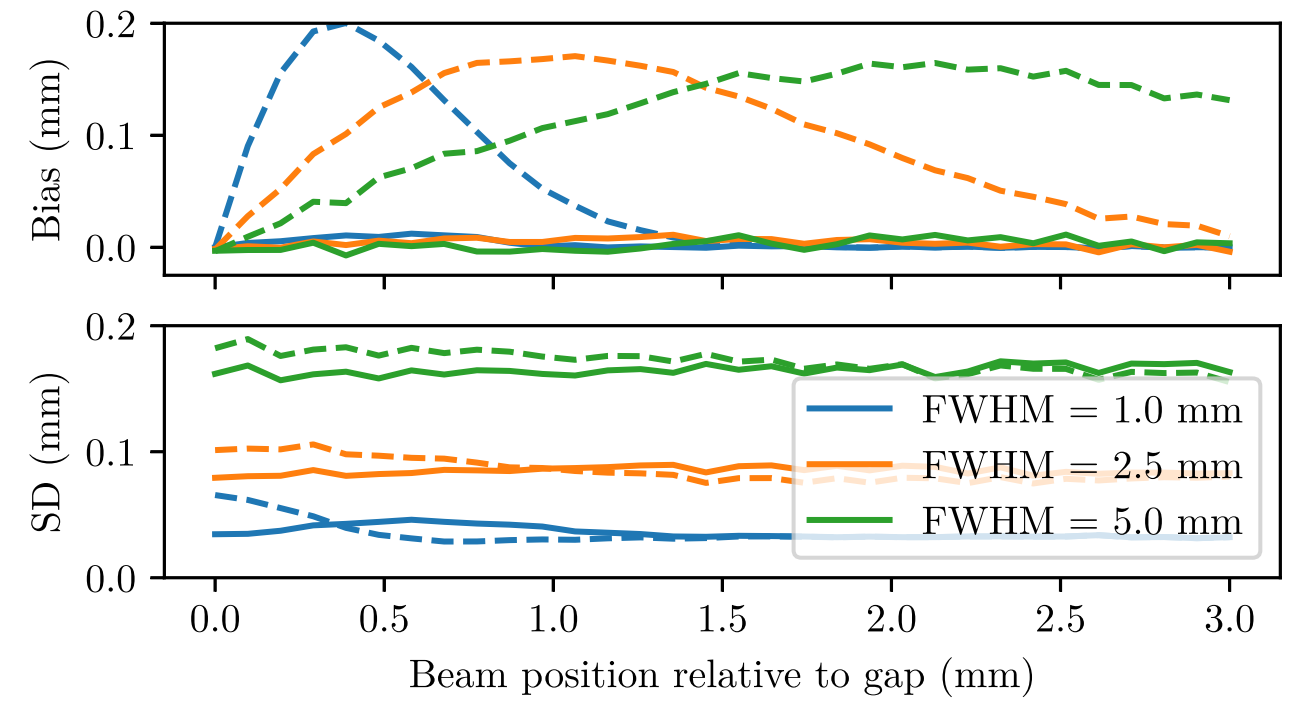}
    \caption{Reconstructed beam position bias (top) and standard deviation (bottom) of simulated frames as a function of the distance to a \SI{600}{\micro\meter} wide insensitive gap from~\cite{pittermann_EvaluationHVCMOSSensors_2022}. Two algorithms are used: simple mean (dashed line) and likelihood fit of a Gaussian function (solid line). The beam is simulated by a random number of hits per frame following Poisson statistics and a Gaussian distribution. The true beam position is varied and three beam widths (different colors) are tested.}
    \label{fig:sim_gap}
\end{figure}

\begin{figure}
    \centering
    \includegraphics[width=.8\textwidth]{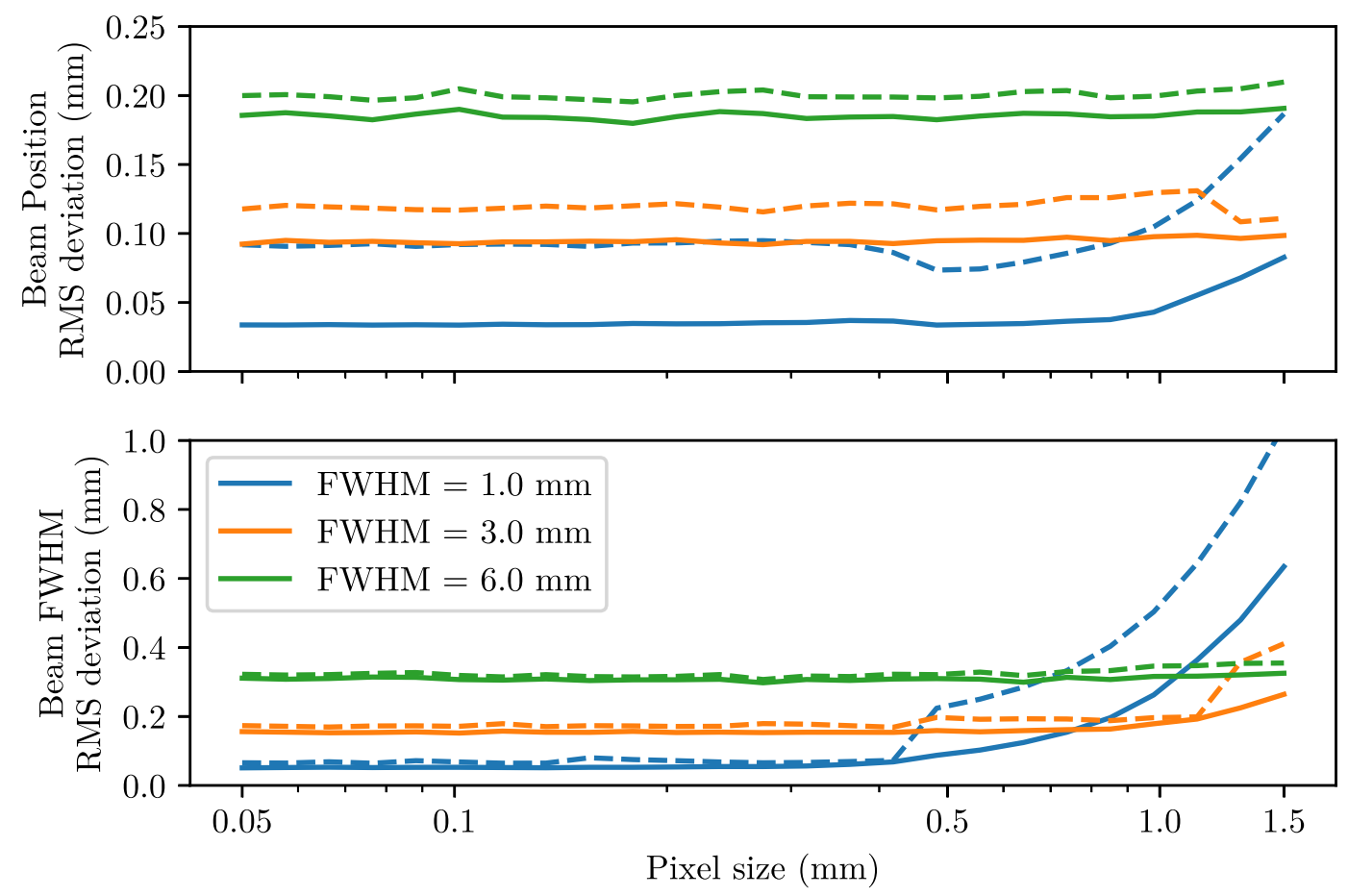}
    \caption{Uncertainty (RMS deviation) of reconstructed beam position (top) and beam width (bottom) of simulated frames as a function of pixel size~\cite{pittermann_EvaluationHVCMOSSensors_2022}. Two algorithms are used: simple mean (dashed line) and likelihood fit of a Gaussian function (solid line). The beam is simulated by a random number of hits per frame following Poisson statistics and a Gaussian distribution. The pixel size is varied and three beam widths (different colors) are tested.}
    \label{fig:sim_pixel}
\end{figure}

Since the proposed beam monitoring system marks a transition from gas filled chambers to solid state detectors, the material effect on the beam needs to be studied.
For this, a simulation with the GEANT4 toolkit~\cite{allison_RecentDevelopmentsGeant4_2016} has been performed comparing the current system at HIT with the proposed system described in Section~\ref{sec:proposed_system}.
The setup of the simulation is illustrated in Figure~\ref{fig:sim_scene}.
The narrow blue layer represents the three ICs with a water equivalent thickness of $3\times$\SI{230}{\micro\meter}~\cite{parodi_MonteCarloSimulations_2012a}.
This corresponds to the scenario in which the HitPix system only replaces the MWCs for beam position and shape monitoring.
In another scenario, the HitPix system could even replace the ICs if the dose measurement is proven to be compatible.
This would reduce the material budget further.
The green layers in Figure~\ref{fig:sim_scene} are either the MWCs (\SI{160}{\micro\meter} water equivalent~\cite{parodi_MonteCarloSimulations_2012a}) or the HitPix system.
The HitPix system consists of \SI{200}{\micro\meter} carbon-fiber-reinforced polymer as the thermal and mechanical support, \SI{100}{\micro\meter} of silicon for the thinned detectors and \SI{90}{\micro\meter} polyimide as flex cable with \SI{7}{\micro\meter} of either aluminum or copper per side.
An area factor is applied for the metal traces to account for \SI{80}{\micro\meter} wide traces at \SI{200}{\micro\meter} pitch.
For comparison, carbon ions at \SI{128.1}{\mega\electronvolt\per u} were used as primary beam traversing the beam monitoring arrangement and entering a water target after about \SI{1.2}{\meter} in air.
The spatial distribution of the energy deposition was recorded.
The resulting profiles are compared in Figure~\ref{fig:scenes_compare} for the current system at HIT, a water equivalent of \SI{2}{\milli\meter}, which is the specified maximum material for five layers (Section~\ref{sec:BAMS}), and the HitPix system in two variants: with a single-layer flex cable made of aluminum traces, and with a double-layer flex cable made of copper traces.
The distributions do not show significant deviations.
As expected, the current system with gas chambers does not affect the low energy carbon beam as much as the solid state approach.
Still, the additional broadening of the beam is not very large if one avoids copper traces.
\begin{table}[]
    \caption{Simulation results of the widening of a carbon ion beam after passing different detector arrangements and entering a water target. The width of the energy deposition is extracted from Figure~\ref{fig:scenes_compare} by fitting two Gaussian functions. Beam parameters: carbon ions, \SI{128.1}{\mega\electronvolt\per u}, parallel beam, point-like beam spot.}
    \centering
    \begin{tabular}{c|c}
        \toprule
        Arrangement & Energy deposition \\ 
         & FWHM width \\ \midrule
        \texttt{HIT current} (2 $\times$ MWC + 3 $\times$ IC) & \SI{3.03}{\milli\meter}\\
        \texttt{\SI{2}{\milli\meter} WE} (water equivalent) & \SI{3.60}{\milli\meter}\\
        \texttt{HitPix (Al single)} (2 $\times$ HitPix  + 3 $\times$ IC) & \SI{3.79}{\milli\meter}\\
        \texttt{HitPix (Cu double)} (2 $\times$ HitPix  + 3 $\times$ IC) & \SI{4.23}{\milli\meter}\\
    \end{tabular}
    \label{tab:sim_results}
\end{table}
Table~\ref{tab:sim_results} summarizes the extracted FWHM for the four scenarios.
At low energy, envisaged arrangement \texttt{HitPix (Al single)} broadens the carbon ion beam by about 5\% more than the \SI{2}{\milli\meter} water equivalent, which represents a tolerable material budget.
The almost homogeneous material distribution of the proposed detector is an advantage compared to the tungsten wires in MWCs, which act as strong scattering centers.\\
The HitPix detectors have a very small periphery to cover a maximally sensitive area.
Still, manufacturing aspects cannot avoid a small insensitive region at three edges of the detector of \SI{150}{\micro\meter}, and at the forth side the end-of-column electronics (\SI{128}{\micro\meter}) together with the contact pad region (\SI{260}{\micro\meter}) add up to a width of about \SI{400}{\micro\meter}.
Also the side-by-side placement of the detectors adds a small gap of about \SI{100}{\micro\meter}.
In summary, there can be insensitive gaps of up to \SI{500}{\micro\meter} every \SI{20}{\milli\meter}, which is the final detector chip dimension.
To study the effect of a worst case gap of \SI{600}{\micro\meter} (equivalent to three missing pixel columns) on the beam spot reconstruction, a simulation of the hit distribution for various Gaussian beam shapes and positions has been performed.
Figure~\ref{fig:sim_gap_illustration} shows two examples of the beam shapes with the gap at around position 2.5.
Distributions following such shapes are generated by Poisson random numbers for an average intensity of \SI{2e6}{\per\second} and \SI{100}{\micro\second} integration time.
In Figure~\ref{fig:sim_gap}, the mean and standard deviation of the reconstructed position from \num{1500} frames is plotted in dependence of the beam distance to the gap applying two different reconstruction algorithms.
The simple mean algorithm is visualized by the dashed lines, while the solid lines represent the results of a likelihood fit with a Gaussian function.
The upper plot shows the average deviation (bias) which is the offset of the reconstructed position compared to the true position.
It is mainly unaffected for the likelihood fit, while the simple mean algorithm has a maximum bias of close to \SI{200}{\micro\meter} when the gap lies in the beam spot tails breaking the symmetry.
A systematic shift in the position information for the feedback loop has to be avoided otherwise the uniform irradiation may be degraded.
The standard deviation (SD) of the beam spot position in the lower plot of Figure~\ref{fig:sim_gap} shows no significant difference between the algorithms.
Wider beams result in a worse standard deviation or uncertainty due to the reduced hit counts per pixel.\\
The appropriate dimensioning of the pixels to mainly record single pixel clusters was validated by measuring the cluster size in Section~\ref{sec:hit_detection}. 
With the simulation framework introduced above, the effect of the pixel dimensions on the precision of the reconstructed beam position and width has been investigated.
The results are shown on Figure~\ref{fig:sim_pixel} for different beam widths, pixel sizes and algorithms.
These plots demonstrate that pixel sizes below \SI{400}{\um} result in similar uncertainties even for beams as narrow as \SI{1}{\mm} and can be implemented in the detector.
Therefore, the pixel size can be adapted to other constraints like readout speed, diode capacitance or power consumption.
%%%%%%%%%%%%%%%%%%%%%%%%%%%%%%%%%%%%%%%%%%
\section{Summary and Conclusion}
This article describes a new detector system based on hit counting HV-CMOS detectors which targets the use as a beam monitor for ion therapy.
The detector is tailored to the requirements of the beam monitoring system at the Heidelberg Ion beam Therapy center.\\
The pixel size is adapted to the resolution requirements.
The hit counting capability and the built-in projections aim for processing high particle rates and fast readout at minimum latency.
Attention was put into avoiding inactive area and achieving low power consumption.\\
In several studies, the properties of the detector in view of the requirements as beam monitor have been investigated.
The resolution of the beam position and spot width for small beam spots were well within the requirements of \SI{200}{\micro\meter} accuracy.
Hits from the particle beam can be recorded continuously with an integration time of \SI{20}{\micro\second} and the scanned beam can be traced.\\
The HV-CMOS detectors were irradiated to a fluence corresponding to the maximum expected fluence in the center of the beam monitor after one year.
The irradiated samples did work in a carbon ion beam with reduced hit detection efficiency which could be corrected for after calibration.
There are options to improve the performance like optimizing the threshold, cooling the detectors and operating at higher bias voltage, which could extent the period for calibrations.\\
Measurements with high-rate ion beams revealed a hit rate limiting effect, which is being investigated.
A likely cause are heavily ionizing particles generating and accumulating charge below the cell electronics which cause the baseline voltage to drop, making following hits undetectable. 
This failure mode will be further analyzed and corrected in the next version.
One possible correction would be to modify the feedback circuit to be able to restore the baseline faster.\\
The detectors were also tested in magnetic fields and demonstrated tolerance of static magnetic fields of at least \SI{250}{\milli\tesla}.
The rapidly changing fields of a scanning MRI induced high voltages in the connecting cables, which were not optimized for this application yet.
In some scanning modes, these voltage pulses on the configuration lines made operation in a scanning MRI system impossible.
The next version of the detector chip will get lockable configuration registers.\\
A first step towards a large matrix of HitPix detectors was the production and partial assembly of a PCB providing supply and data lines for up to ten chips.
The daisy chain and bus readout modes were tested successfully and the matrix recorded particle beams at HIT.
The reconstructed beam spot widths are well in agreement with the nominal beam settings.\\
In addition to measurements with the detectors, several simulation studies were performed.
The impact of the solid-state detector on the scattering of the beam is close to the specified upper limit of \SI{2}{\milli\meter} water equivalent, and has the advantage that concentrated scattering centers, like the wires of the multi-wire chambers, are not present. 
The effect on the position resolution with unavoidable gaps of sensitive area between detector chips was studied.
In a worst-case scenario with \SI{600}{\micro\meter} gaps and a very narrow beam of \SI{1}{\milli\meter} (FWHM), the bias of the position reaches \SI{200}{\micro\meter} at the worst position about \SI{400}{\micro\meter} away from the gap.
This is true for a simple algorithm calculating the mean of the hit distribution.
A likelihood fit of a Gaussian function would not be affected at all by the gaps and will be considered for implementation on an FPGA.
In another simulation, it was shown that pixel sizes up to \SI{400}{\um} show similar and adequate reconstruction accuracy of the beam spot with a width down to \SI{1}{\mm}.

In summary, the investigated HitPix2 detector has proven to be a promising candidate for a beam monitoring system at an ion beam facility.
The shortcomings identified during laboratory and beam tests are being addressed and a successor is in development.
The next HitPix version will feature offset tuning, the second projection direction, faster readout concept, and increased tolerance against single-event and charge-up effects.
The next step is to assemble a matrix as envisioned in Section~\ref{sec:proposed_system} but with a reduced number of $5 \times 5$ chips.
This detector system will be used for long-term operation tests in the QA room at HIT.
\newpage
%%%%%%%%%%%%%%%%%%%%%%%%%%%%%%%%%%%%%%%%%%
\authorcontributions{Conceptualization, Alexander Dierlamm, Ivan Perić, Alena Weber and Jürgen Debus; Data curation, Martin Pittermann; Formal analysis, Martin Pittermann, Bogdan Topko, Thomas Hansmann and Jakob Naumann; Funding acquisition, Alexander Dierlamm, Felix Ehrler, Ulrich Husemann, Ivan Perić, Jürgen Debus, Nicole Grau, Oliver Jäkel and Sebastian Kluter; Investigation, Felix Ehrler, Roland Koppenhöfer, Martin Pittermann, Bogdan Topko, Alena Weber and Jakob Naumann; Project administration, Alexander Dierlamm and Ulrich Husemann; Resources, Felix Ehrler, Ivan Perić, Martin Pittermann, Alena Weber, Stephan Brons and Jakob Naumann; Software, Felix Ehrler and Martin Pittermann; Supervision, Matthias Balzer, Ulrich Husemann, Ivan Perić and Sebastian Kluter; Validation, Alexander Dierlamm and Jakob Naumann; Visualization, Felix Ehrler, Martin Pittermann and Bogdan Topko; Writing – original draft, Alexander Dierlamm and Felix Ehrler; Writing – review and editing, Ulrich Husemann, Roland Koppenhöfer, Nicole Grau, Thomas Hansmann, Oliver Jäkel and Jakob Naumann.}

\funding{This research was funded by the Heidelberg Karlsruhe Strategic Partnership (HEiKA), the KIT Center Elementary Particle and Astroparticle Physics (KCETA), Helmholtz Program Matter and Technology, German Federal Ministry of Education (BMBF) within the Adaptive RadioThErapie Mit IonenStrahlen (ARTEMIS) Project under Grant 13GW0436A.}

\acknowledgments{The authors would like to thank Rudolf Schimassek for his contribution to the DAQ system and GECCO board, Horacio Mateos for the assembly and bonding and Bernd Berger for performing the irradiations at the cyclotron.}

%%%%%%%%%%%%%%%%%%%%%%%%%%%%%%%%%%%%%%%%%%
\begin{adjustwidth}{-\extralength}{0cm}

\reftitle{References}

\bibliography{HitPix}

\begin{thebibliography}{999}

\bibitem[J{\"a}kel \em{et~al.}(2022)J{\"a}kel, Kraft, and
  Karger]{jakel_HistoryIonBeam_2022}
J{\"a}kel, O.; Kraft, G.; Karger, C.P.
\newblock The History of Ion Beam Therapy in {{Germany}}.
\newblock {\em Zeitschrift f\"ur Medizinische Physik} {\bf 2022}, {\em
  32},~6--22.
\newblock {\url{https://doi.org/10.1016/j.zemedi.2021.11.003}}.

\bibitem[Royce and Efstathiou(2019)]{royce_ProtonTherapyProstate_2019}
Royce, T.J.; Efstathiou, J.A.
\newblock Proton Therapy for Prostate Cancer: {{A}} Review of the Rationale,
  Evidence, and Current State.
\newblock {\em Urologic Oncology: Seminars and Original Investigations} {\bf
  2019}, {\em 37},~628--636.
\newblock {\url{https://doi.org/10.1016/j.urolonc.2018.11.012}}.

\bibitem[Haberer \em{et~al.}(2004)Haberer, Debus, Eickhoff, J{\"a}kel,
  {Schulz-Ertner}, and Weber]{haberer_HeidelbergIonTherapy_2004}
Haberer, T.; Debus, J.; Eickhoff, H.; J{\"a}kel, O.; {Schulz-Ertner}, D.;
  Weber, U.
\newblock The Heidelberg Ion Therapy Center.
\newblock {\em Radiotherapy and Oncology} {\bf 2004}, {\em 73},~S186--S190.
\newblock {\url{https://doi.org/10.1016/S0167-8140(04)80046-X}}.

\bibitem[Haberer \em{et~al.}(1993)Haberer, Becher, Schardt, and
  Kraft]{haberer_MagneticScanningSystem_1993}
Haberer, T.; Becher, W.; Schardt, D.; Kraft, G.
\newblock Magnetic Scanning System for Heavy Ion Therapy.
\newblock {\em Nuclear Instruments and Methods in Physics Research Section A:
  Accelerators, Spectrometers, Detectors and Associated Equipment} {\bf 1993},
  {\em 330},~296--305.
\newblock {\url{https://doi.org/10.1016/0168-9002(93)91335-K}}.

\bibitem[Galonska \em{et~al.}(2013)Galonska, Scheloske, Brons, Cee,
  H{\"o}ppner, Mosthaf, Peters, and
  Haberer]{galonska_HitGantryCommissioning_2013}
Galonska, M.; Scheloske, S.; Brons, S.; Cee, R.; H{\"o}ppner, K.; Mosthaf, J.;
  Peters, A.; Haberer, T.
\newblock The Hit Gantry: {{From}} Commissioning to Operation.
\newblock {\em IPAC 2013: Proceedings of the 4th International Particle
  Accelerator Conference} {\bf 2013}, pp. 3636--3638.

\bibitem[Kleffner \em{et~al.}(2009)Kleffner, Ondreka, and
  Weinrich]{kleffner_HeidelbergIonTherapy_2009}
Kleffner, C.; Ondreka, D.; Weinrich, U.
\newblock The {{Heidelberg Ion Therapy}} ({{HIT}}) {{Accelerator Coming}} into
  {{Operation}}.
\newblock {\em AIP Conference Proceedings} {\bf 2009}, {\em 1099},~426--428.
\newblock {\url{https://doi.org/10.1063/1.3120065}}.

\bibitem[Sch{\"o}mers \em{et~al.}(2013)Sch{\"o}mers, Feldmeier, Haberer,
  Naumann, Panse, and
  Peters]{schomers_PatientspecificIntensitymodulationSlowly_2013}
Sch{\"o}mers, C.; Feldmeier, E.; Haberer, T.; Naumann, J.; Panse, R.; Peters,
  A.
\newblock Patient-Specific {{Intensity-modulation}} of a {{Slowly Extracted
  Beam}} at the {{HIT Synchrotron}}.
\newblock In Proceedings of the {{IPAC2013}}; ,  2013; pp. 2944--2946.

\bibitem[Kl{\"u}ter(2019)]{kluter_TechnicalDesignConcept_2019}
Kl{\"u}ter, S.
\newblock Technical Design and Concept of a 0.35 {{T MR-Linac}}.
\newblock {\em Clinical and Translational Radiation Oncology} {\bf 2019}, {\em
  18},~98--101.
\newblock {\url{https://doi.org/10.1016/j.ctro.2019.04.007}}.

\bibitem[Hoffmann \em{et~al.}(2020)Hoffmann, Oborn, Moteabbed, Yan, Bortfeld,
  Knopf, Fuchs, Georg, Seco, Spadea, J{\"a}kel, Kurz, and
  Parodi]{hoffmann_MRguidedProtonTherapy_2020}
Hoffmann, A.; Oborn, B.; Moteabbed, M.; Yan, S.; Bortfeld, T.; Knopf, A.;
  Fuchs, H.; Georg, D.; Seco, J.; Spadea, M.F.;  et~al.
\newblock {{MR-guided}} Proton Therapy: A Review and a Preview.
\newblock {\em Radiation Oncology (London, England)} {\bf 2020}, {\em 15},~129.
\newblock {\url{https://doi.org/10.1186/s13014-020-01571-x}}.

\bibitem[Sauli(1977)]{sauli_PrinciplesOperationMultiwire_1977}
Sauli, F.
\newblock Principles of Operation of Multiwire Proportional and Drift Chambers.
\newblock {\em CERN Yellow Report} {\bf 1977}.
\newblock {\url{https://doi.org/10.5170/CERN-1977-009}}.

\bibitem[Lin \em{et~al.}(2009)Lin, Boehringer, Coray, Grossmann, and
  Pedroni]{lin_More10Years_2009}
Lin, S.; Boehringer, T.; Coray, A.; Grossmann, M.; Pedroni, E.
\newblock More than 10 Years Experience of Beam Monitoring with the {{Gantry}}
  1 Spot Scanning Proton Therapy Facility at {{PSI}}.
\newblock {\em Medical Physics} {\bf 2009}, {\em 36},~5331--5340.
\newblock {\url{https://doi.org/10.1118/1.3244034}}.

\bibitem[Giordanengo \em{et~al.}(2017)Giordanengo, Manganaro, and
  Vignati]{giordanengo_ReviewTechnologiesProcedures_2017}
Giordanengo, S.; Manganaro, L.; Vignati, A.
\newblock Review of Technologies and Procedures of Clinical Dosimetry for
  Scanned Ion Beam Radiotherapy.
\newblock {\em Physica Medica} {\bf 2017}, {\em 43},~79--99.
\newblock {\url{https://doi.org/10.1016/j.ejmp.2017.10.013}}.

\bibitem[Wang \em{et~al.}(2017)Wang, Zou, Fan, Liu, Sun, Wang, Kang, Sun, Yang,
  Pei, Huang, Xu, Gao, and Xiao]{wang_BeamMonitorUsing_2017}
Wang, Z.; Zou, S.; Fan, Y.; Liu, J.; Sun, X.; Wang, D.; Kang, H.; Sun, D.;
  Yang, P.; Pei, H.;  et~al.
\newblock A Beam Monitor Using Silicon Pixel Sensors for Hadron Therapy.
\newblock {\em Nuclear Instruments and Methods in Physics Research Section A:
  Accelerators, Spectrometers, Detectors and Associated Equipment} {\bf 2017},
  {\em 849},~20--24.
\newblock {\url{https://doi.org/10.1016/j.nima.2016.12.050}}.

\bibitem[Actis \em{et~al.}(2014)Actis, Meer, and
  K{\"o}nig]{actis_PreciseOnlinePosition_2014}
Actis, O.; Meer, D.; K{\"o}nig, S.
\newblock Precise On-Line Position Measurement for Particle Therapy.
\newblock {\em Journal of Instrumentation} {\bf 2014}, {\em 9},~C12037--C12037.
\newblock {\url{https://doi.org/10.1088/1748-0221/9/12/C12037}}.

\bibitem[Klimpki \em{et~al.}(2017)Klimpki, Psoroulas, Bula, Rechsteiner,
  Eichin, Weber, Lomax, and Meer]{klimpki_BeamMonitoringValidation_2017}
Klimpki, G.; Psoroulas, S.; Bula, C.; Rechsteiner, U.; Eichin, M.; Weber, D.C.;
  Lomax, A.; Meer, D.
\newblock A Beam Monitoring and Validation System for Continuous Line Scanning
  in Proton Therapy.
\newblock {\em Physics in Medicine \& Biology} {\bf 2017}, {\em
  62},~6126--6143.
\newblock {\url{https://doi.org/10.1088/1361-6560/aa772e}}.

\bibitem[Papa(2016)]{papa_ScintillatingFibresCoupled_2016}
Papa, A.
\newblock Scintillating Fibres Coupled to Silicon Photomultiplier Prototypes
  for Fast Beam Monitoring and Thin Timing Detectors.
\newblock {\em Nuclear Instruments and Methods in Physics Research Section A:
  Accelerators, Spectrometers, Detectors and Associated Equipment} {\bf 2016},
  {\em 824},~128--130.
\newblock {\url{https://doi.org/10.1016/j.nima.2015.11.097}}.

\bibitem[Ortega~Ruiz \em{et~al.}(2019)Ortega~Ruiz, Fosse, Franchi, Frassier,
  Fullerton, Kral, Lauener, Schneider, Spanggaard, and
  Tranquille]{ortegaruiz_MultipurposeScintillatingFibre_2019a}
Ortega~Ruiz, I.; Fosse, L.; Franchi, J.; Frassier, A.; Fullerton, J.; Kral, J.;
  Lauener, J.; Schneider, T.; Spanggaard, J.; Tranquille, G.
\newblock A {{Multipurpose Scintillating Fibre Beam Monitor}} for the
  {{Measurement}} of {{Secondary Beams}} at {{CERN}}.
\newblock {\em Proceedings of the 7\textbackslash textsuperscript\{th\} Int.
  Beam Instrumentation Conf.} {\bf 2019}, {\em IBIC2018},~4.
\newblock {\url{https://doi.org/10.18429/JACOW-IBIC2018-WEPB15}}.

\bibitem[Allegrini \em{et~al.}(2021)Allegrini, Cachemiche, Caplan, Carlus,
  Chen, Curtoni, Dauvergne, Negra, {Gallin-Martel}, H{\'e}rault, L{\'e}tang,
  Morel, Testa, and
  Zoccarato]{allegrini_CharacterizationBeamtaggingHodoscope_2021}
Allegrini, O.; Cachemiche, J.P.; Caplan, C.P.C.; Carlus, B.; Chen, X.; Curtoni,
  S.; Dauvergne, D.; Negra, R.D.; {Gallin-Martel}, M.L.; H{\'e}rault, J.;
  et~al.
\newblock Characterization of a Beam-Tagging Hodoscope for Hadrontherapy
  Monitoring.
\newblock {\em Journal of Instrumentation} {\bf 2021}, {\em
  16},~P02028--P02028.
\newblock {\url{https://doi.org/10.1088/1748-0221/16/02/P02028}}.

\bibitem[Joram \em{et~al.}(2015)Joram, Uwer, Kirn, Leverington, Bachmann,
  Ekelhof, and M{\"u}ller]{joram_LHCbScintillatingFibre_2015}
Joram, C.; Uwer, U.; Kirn, T.; Leverington, B.D.; Bachmann, S.; Ekelhof, R.J.;
  M{\"u}ller, J.
\newblock {LHCb Scintillating Fibre Tracker Engineering Design Review Report:
  Fibres, Mats and Modules}.
\newblock Technical Report LHCb-PUB-2015-008, {CERN},  2015.

\bibitem[Leverington \em{et~al.}(2018)Leverington, Dziewiecki, Renner, and
  Runze]{leverington_PrototypeScintillatingFibre_2018}
Leverington, B.; Dziewiecki, M.; Renner, L.; Runze, R.
\newblock A Prototype Scintillating Fibre Beam Profile Monitor for {{Ion
  Therapy}} Beams.
\newblock {\em Journal of Instrumentation} {\bf 2018}, {\em
  13},~P05030--P05030.
\newblock {\url{https://doi.org/10.1088/1748-0221/13/05/P05030}}.

\bibitem[Marti{\v s}{\'i}kov{\'a} \em{et~al.}(2011)Marti{\v s}{\'i}kov{\'a},
  Hesse, Nairz, and J{\"a}kel]{martisikova_TestAmorphousSilicon_2011}
Marti{\v s}{\'i}kov{\'a}, M.; Hesse, B.M.; Nairz, O.; J{\"a}kel, O.
\newblock Test of an Amorphous Silicon Detector in Medical Proton Beams.
\newblock {\em Nuclear Instruments and Methods in Physics Research Section A:
  Accelerators, Spectrometers, Detectors and Associated Equipment} {\bf 2011},
  {\em 633},~S259--S261.
\newblock {\url{https://doi.org/10.1016/j.nima.2010.06.182}}.

\bibitem[Sacchi \em{et~al.}(2020)Sacchi, Ganjeh, Arcidiacono, Attili,
  Cartiglia, Donetti, Fausti, Ferrero, Giordanengo, Ali, Mandurrino, Manganaro,
  Mazza, Monaco, Sola, Staiano, Vignati, and
  Cirio]{sacchi_TestInnovativeSilicon_2020}
Sacchi, R.; Ganjeh, Z.A.; Arcidiacono, R.; Attili, A.; Cartiglia, N.; Donetti,
  M.; Fausti, F.; Ferrero, M.; Giordanengo, S.; Ali, O.H.;  et~al.
\newblock Test of Innovative Silicon Detectors for the Monitoring of a
  Therapeutic Proton Beam.
\newblock {\em Journal of Physics: Conference Series} {\bf 2020}, {\em
  1662},~012002.
\newblock {\url{https://doi.org/10.1088/1742-6596/1662/1/012002}}.

\bibitem[Vignati \em{et~al.}(2020)Vignati, Donetti, Fausti, Ferrero,
  Giordanengo, Ali, Villarreal, Milian, Mazza, Monaco, Sacchi, Shakarami, Sola,
  Staiano, Tommasino, Verroi, Wheadon, and
  Cirio]{vignati_ThinLowgainAvalanche_2020}
Vignati, A.; Donetti, M.; Fausti, F.; Ferrero, M.; Giordanengo, S.; Ali, O.H.;
  Villarreal, O.A.M.; Milian, F.M.; Mazza, G.; Monaco, V.;  et~al.
\newblock Thin Low-Gain Avalanche Detectors for Particle Therapy Applications.
\newblock {\em Journal of Physics: Conference Series} {\bf 2020}, {\em
  1662},~012035.
\newblock {\url{https://doi.org/10.1088/1742-6596/1662/1/012035}}.

\bibitem[Kr{\"u}ger \em{et~al.}(2022)Kr{\"u}ger, Bergauer, Galatyuk, Hirtl,
  Kedych, Kis, Linev, Michel, Pietraszko, Pitters, Rost, Schmidt,
  Svintozelskyi, Tr{\"a}ger, Traxler, {Ulrich-Pur}, and
  Wendisch]{kruger_LGADTechnologyHADES_2022}
Kr{\"u}ger, W.; Bergauer, T.; Galatyuk, T.; Hirtl, A.; Kedych, V.; Kis, M.;
  Linev, S.; Michel, J.; Pietraszko, J.; Pitters, F.;  et~al.
\newblock {{LGAD}} Technology for {{HADES}}, Accelerator and Medical
  Applications.
\newblock {\em Nuclear Instruments and Methods in Physics Research Section A:
  Accelerators, Spectrometers, Detectors and Associated Equipment} {\bf 2022},
  {\em 1039},~167046.
\newblock {\url{https://doi.org/10.1016/j.nima.2022.167046}}.

\bibitem[Flynn \em{et~al.}(2022)Flynn, Manolopoulos, Rompokos, Poynter, Toltz,
  Beck, Ballisat, Velthuis, Allport, Green, Thomas, and
  Price]{flynn_MonitoringPencilBeam_2022}
Flynn, S.; Manolopoulos, S.; Rompokos, V.; Poynter, A.; Toltz, A.; Beck, L.;
  Ballisat, L.; Velthuis, J.; Allport, P.; Green, S.;  et~al.
\newblock Monitoring Pencil Beam Scanned Proton Radiotherapy Using a Large
  Format {{CMOS}} Detector.
\newblock {\em Nuclear Instruments and Methods in Physics Research Section A:
  Accelerators, Spectrometers, Detectors and Associated Equipment} {\bf 2022},
  {\em 1033},~166703.
\newblock {\url{https://doi.org/10.1016/j.nima.2022.166703}}.

\bibitem[Hartmann(2011)]{hartmann_SemiconductorSensors_2011}
Hartmann, F.
\newblock Semiconductor Sensors.
\newblock {\em Nuclear Instruments and Methods in Physics Research Section A:
  Accelerators, Spectrometers, Detectors and Associated Equipment} {\bf 2011},
  {\em 628},~40--49.
\newblock {\url{https://doi.org/10.1016/j.nima.2010.06.282}}.

\bibitem[Bruzzi(2016)]{bruzzi_NovelSiliconDevices_2016}
Bruzzi, M.
\newblock Novel {{Silicon Devices}} for {{Radiation Therapy Monitoring}}.
\newblock {\em Nuclear Instruments and Methods in Physics Research Section A:
  Accelerators, Spectrometers, Detectors and Associated Equipment} {\bf 2016},
  {\em 809},~105--112.
\newblock {\url{https://doi.org/10.1016/j.nima.2015.10.072}}.

\bibitem[Hartmann(2017)]{hartmann_EvolutionSiliconSensor_2017}
Hartmann, F.
\newblock {\em Evolution of Silicon Sensor Technology in Particle Physics};
  {Springer Berlin Heidelberg}: {New York, NY},  2017.

\bibitem[Lutz(2007)]{lutz_SemiconductorRadiationDetectors_2007}
Lutz, G.
\newblock {\em Semiconductor Radiation Detectors: Device Physics}; {Springer}:
  {Berlin},  2007.

\bibitem[Peri{\'c} \em{et~al.}(2011)Peri{\'c}, Kreidl, and
  Fischer]{peric_ParticlePixelDetectors_2011}
Peri{\'c}, I.; Kreidl, C.; Fischer, P.
\newblock Particle Pixel Detectors in High-Voltage {{CMOS}}
  Technology\textemdash{{New}} Achievements.
\newblock {\em Nuclear Instruments and Methods in Physics Research Section A:
  Accelerators, Spectrometers, Detectors and Associated Equipment} {\bf 2011},
  {\em 650},~158--162.
\newblock {\url{https://doi.org/10.1016/j.nima.2010.11.090}}.

\bibitem[Moll(1999)]{moll_RadiationDamageSilicon_1999}
Moll, M.
\newblock Radiation Damage in Silicon Particle Detectors: Microscopic Defects
  and Macroscopic Properties.
\newblock PhD thesis, University of Hamburg,  1999.

\bibitem[Arndt \em{et~al.}(2021)Arndt, Augustin, Baesso, Berger, Berg,
  Betancourt, Bortoletto, Bravar, Briggl, {vom Bruch}, Buonaura, Cadoux,
  Chavez~Barajas, Chen, Clark, Cooke, Corrodi, Damyanova, Demets, Dittmeier,
  Eckert, Ehrler, Fahrni, Gagneur, Gerritzen, Goldstein, Gottschalk, Grab,
  Gredig, Groves, Hammerich, Hartenstein, Hartmann, Hayward, Herkert, Hesketh,
  Hetzel, Hildebrandt, Hodge, Hofer, Huang, Hughes, Huth, Immig, Jones, Jones,
  K{\"a}stli, K{\"o}ppel, Kettle, Kiehn, Kilani, Klingenmeyer, Knecht, Knight,
  Kotlinski, Kozlinskiy, Leys, Lockwood, Loreti, La~Marra, M{\"u}ller, Meier,
  Meier~Aeschbacher, Meneses, Metodiev, Mtchedlishvili, Muley, Munwes, Noehte,
  Owen, Papa, Paraskevas, Peri{\'c}, Perrevoort, Plackett, Pohl, Ritt, Robmann,
  Rompotis, Rudzki, Rutar, Sch{\"o}ning, Schimassek, {Schultz-Coulon}, Serra,
  Shen, Shipsey, Shrestha, Steinkamp, Stoykov, Straumann, Streuli, Stumpf,
  Tata, Velthuis, Vigani, {Vilella-Figueras}, Vossebeld, Wallny, Wasili,
  Wauters, Weber, Wiedner, Windelband, and
  Zhong]{arndt_TechnicalDesignPhase_2021}
Arndt, K.; Augustin, H.; Baesso, P.; Berger, N.; Berg, F.; Betancourt, C.;
  Bortoletto, D.; Bravar, A.; Briggl, K.; {vom Bruch}, D.;  et~al.
\newblock Technical Design of the Phase {{I Mu3e}} Experiment.
\newblock {\em Nuclear Instruments and Methods in Physics Research Section A:
  Accelerators, Spectrometers, Detectors and Associated Equipment} {\bf 2021},
  {\em 1014},~165679.
\newblock {\url{https://doi.org/10.1016/j.nima.2021.165679}}.

\bibitem[Augustin \em{et~al.}(2020)Augustin, Peri{\'c}, Sch{\"o}ning, and
  Weber]{augustin_MuPixSensorMu3e_2020}
Augustin, H.; Peri{\'c}, I.; Sch{\"o}ning, A.; Weber, A.
\newblock The {{MuPix}} Sensor for the {{Mu3e}} Experiment.
\newblock {\em Nuclear Instruments and Methods in Physics Research Section A:
  Accelerators, Spectrometers, Detectors and Associated Equipment} {\bf 2020},
  {\em 979},~164441.
\newblock {\url{https://doi.org/10.1016/j.nima.2020.164441}}.

\bibitem[Schimassek \em{et~al.}(2021)Schimassek, Andreazza, Augustin, Barbero,
  Benoit, Ehrler, Iacobucci, Meneses, Pangaud, Prathapan, Sch{\"o}ning,
  Vilella, Weber, Weber, Wong, Zhang, and
  Peri{\'c}]{schimassek_TestResultsATLASPIX3_2021}
Schimassek, R.; Andreazza, A.; Augustin, H.; Barbero, M.; Benoit, M.; Ehrler,
  F.; Iacobucci, G.; Meneses, A.; Pangaud, P.; Prathapan, M.;  et~al.
\newblock Test Results of {{ATLASPIX3}} \textemdash{} {{A}} Reticle Size
  {{HVCMOS}} Pixel Sensor Designed for Construction of Multi Chip Modules.
\newblock {\em Nuclear Instruments and Methods in Physics Research Section A:
  Accelerators, Spectrometers, Detectors and Associated Equipment} {\bf 2021},
  {\em 986},~164812.
\newblock {\url{https://doi.org/10.1016/j.nima.2020.164812}}.

\bibitem[Peri{\'c} \em{et~al.}(2021)Peri{\'c}, Andreazza, Augustin, Barbero,
  Benoit, Casanova, Ehrler, Iacobucci, Leys, Gonzalez, Pangaud, Prathapan,
  Schimassek, Sch{\"o}ning, Figueras, Weber, Weber, Wong, and
  Zhang]{peric_HighVoltageCMOSActive_2021}
Peri{\'c}, I.; Andreazza, A.; Augustin, H.; Barbero, M.; Benoit, M.; Casanova,
  R.; Ehrler, F.; Iacobucci, G.; Leys, R.; Gonzalez, A.M.;  et~al.
\newblock High-{{Voltage CMOS Active Pixel Sensor}}.
\newblock {\em IEEE Journal of Solid-State Circuits} {\bf 2021}, {\em
  56},~2488--2502.
\newblock {\url{https://doi.org/10.1109/JSSC.2021.3061760}}.

\bibitem[Weber \em{et~al.}(2022)Weber, Ehrler, Schimassek, and
  Peri{\'c}]{weber_HighVoltageCMOS_2022}
Weber, A.; Ehrler, F.; Schimassek, R.; Peri{\'c}, I.
\newblock High {{Voltage CMOS}} Active Pixel Sensor Chip with Counting
  Electronics for Beam Monitoring.
\newblock {\em IEEE Transactions on Nuclear Science} {\bf 2022}, {\em
  69},~1288--1298.
\newblock {\url{https://doi.org/10.1109/TNS.2022.3173807}}.

\bibitem[Weber(2021)]{weber_DevelopmentIntegratedCircuits_2021}
Weber, A.L.
\newblock {Development of Integrated Circuits and Smart Sensors for Particle
  Detection in Physics Experiments and Particle Therapy}.
\newblock PhD thesis, University of Heidelberg,  2021.

\bibitem[Schimassek(2021)]{schimassek_EntwicklungUndCharakterisierung_2021}
Schimassek, R.
\newblock {Entwicklung und Charakterisierung von Integrierten Sensoren f\"ur
  die Teilchenphysik = Development and Characterisation of Integrated Sensors
  for Particle Physics}.
\newblock PhD thesis, Karlsruhe Institute of Technology (KIT),  2021.
\newblock {\url{https://doi.org/10.5445/IR/1000141412}}.

\bibitem[Lindstr{\"o}m \em{et~al.}(1999)Lindstr{\"o}m, Moll, and
  Fretwurst]{lindstrom_RadiationHardnessSilicon_1999}
Lindstr{\"o}m, G.; Moll, M.; Fretwurst, E.
\newblock Radiation Hardness of Silicon Detectors \textendash{} a Challenge
  from High-Energy Physics.
\newblock {\em Nuclear Instruments and Methods in Physics Research Section A:
  Accelerators, Spectrometers, Detectors and Associated Equipment} {\bf 1999},
  {\em 426},~1--15.
\newblock {\url{https://doi.org/10.1016/S0168-9002(98)01462-4}}.

\bibitem[Ehrler(2021)]{ehrler_CharakterisierungMonolithischenHVCMOSPixelsensoren_2021}
Ehrler, F.M.
\newblock {Charakterisierung von monolithischen HV-CMOS-Pixelsensoren f\"ur
  Teilchenphysikexperimente = Characterization of monolithic HV-CMOS pixel
  sensors for particle physics experiments}.
\newblock PhD thesis, Karlsruhe Institute of Technology (KIT),  2021.
\newblock {\url{https://doi.org/10.5445/IR/1000133748}}.

\bibitem[Snoeys \em{et~al.}(2000)Snoeys, Faccio, Burns, Campbell, Cantatore,
  Carrer, Casagrande, Cavagnoli, Dachs, Di~Liberto, Formenti, Giraldo, Heijne,
  Jarron, Letheren, Marchioro, Martinengo, Meddi, Mikulec, Morando, Morel,
  Noah, Paccagnella, Ropotar, Saladino, Sansen, Santopietro, Scarlassara,
  Segato, Signe, Soramel, Vannucci, and
  Vleugels]{snoeys_LayoutTechniquesEnhance_2000}
Snoeys, W.; Faccio, F.; Burns, M.; Campbell, M.; Cantatore, E.; Carrer, N.;
  Casagrande, L.; Cavagnoli, A.; Dachs, C.; Di~Liberto, S.;  et~al.
\newblock Layout Techniques to Enhance the Radiation Tolerance of Standard
  {{CMOS}} Technologies Demonstrated on a Pixel Detector Readout Chip.
\newblock {\em Nuclear Instruments and Methods in Physics Research Section A:
  Accelerators, Spectrometers, Detectors and Associated Equipment} {\bf 2000},
  {\em 439},~349--360.
\newblock {\url{https://doi.org/10.1016/S0168-9002(99)00899-2}}.

\bibitem[Mager(2016)]{mager_ALPIDEMonolithicActive_2016}
Mager, M.
\newblock {{ALPIDE}}, the {{Monolithic Active Pixel Sensor}} for the {{ALICE
  ITS}} Upgrade.
\newblock {\em Nuclear Instruments and Methods in Physics Research Section A:
  Accelerators, Spectrometers, Detectors and Associated Equipment} {\bf 2016},
  {\em 824},~434--438.
\newblock {\url{https://doi.org/10.1016/j.nima.2015.09.057}}.

\bibitem[Pittermann(2022)]{pittermann_EvaluationHVCMOSSensors_2022}
Pittermann, M.
\newblock Evaluation of {{HV-CMOS Sensors}} in a {{Beam Monitoring System}} for
  {{Ion Therapy}}.
\newblock Master Thesis ETP-KA/2022-11, {Karlsruhe Institute of Technology},
  2022.

\bibitem[M{\"u}ller(1991)]{muller_GeneralizedDeadTimes_1991}
M{\"u}ller, J.W.
\newblock Generalized Dead Times.
\newblock {\em Nuclear Instruments and Methods in Physics Research Section A:
  Accelerators, Spectrometers, Detectors and Associated Equipment} {\bf 1991},
  {\em 301},~543--551.
\newblock {\url{https://doi.org/10.1016/0168-9002(91)90021-H}}.

\bibitem[Allison \em{et~al.}(2016)Allison, Amako, Apostolakis, Arce, Asai, Aso,
  Bagli, Bagulya, Banerjee, Barrand, Beck, Bogdanov, Brandt, Brown, Burkhardt,
  Canal, {Cano-Ott}, Chauvie, Cho, Cirrone, Cooperman, {Cort{\'e}s-Giraldo},
  Cosmo, Cuttone, Depaola, Desorgher, Dong, Dotti, Elvira, Folger, Francis,
  Galoyan, Garnier, Gayer, Genser, Grichine, Guatelli, Gu{\`e}ye, Gumplinger,
  Howard, H{\v r}ivn{\'a}{\v c}ov{\'a}, Hwang, Incerti, Ivanchenko, Ivanchenko,
  Jones, Jun, Kaitaniemi, Karakatsanis, Karamitros, Kelsey, Kimura, Koi,
  Kurashige, Lechner, Lee, Longo, Maire, Mancusi, Mantero, Mendoza, Morgan,
  Murakami, Nikitina, Pandola, Paprocki, Perl, Petrovi{\'c}, Pia, Pokorski,
  Quesada, Raine, Reis, Ribon, Risti{\'c}~Fira, Romano, Russo, Santin, Sasaki,
  Sawkey, Shin, Strakovsky, Taborda, Tanaka, Tom{\'e}, Toshito, Tran, Truscott,
  Urban, Uzhinsky, Verbeke, Verderi, Wendt, Wenzel, Wright, Wright, Yamashita,
  Yarba, and Yoshida]{allison_RecentDevelopmentsGeant4_2016}
Allison, J.; Amako, K.; Apostolakis, J.; Arce, P.; Asai, M.; Aso, T.; Bagli,
  E.; Bagulya, A.; Banerjee, S.; Barrand, G.;  et~al.
\newblock Recent Developments in {{Geant4}}.
\newblock {\em Nuclear Instruments and Methods in Physics Research Section A:
  Accelerators, Spectrometers, Detectors and Associated Equipment} {\bf 2016},
  {\em 835},~186--225.
\newblock {\url{https://doi.org/10.1016/j.nima.2016.06.125}}.

\bibitem[Parodi \em{et~al.}(2012)Parodi, Mairani, Brons, Hasch, Sommerer,
  Naumann, J{\"a}kel, Haberer, and Debus]{parodi_MonteCarloSimulations_2012a}
Parodi, K.; Mairani, A.; Brons, S.; Hasch, B.G.; Sommerer, F.; Naumann, J.;
  J{\"a}kel, O.; Haberer, T.; Debus, J.
\newblock Monte {{Carlo}} Simulations to Support Start-up and Treatment
  Planning of Scanned Proton and Carbon Ion Therapy at a Synchrotron-Based
  Facility.
\newblock {\em Physics in Medicine \& Biology} {\bf 2012}, {\em 57},~3759.
\newblock {\url{https://doi.org/10.1088/0031-9155/57/12/3759}}.

\end{thebibliography}
% requires (Better) BibTeX export, not BibLaTeX

\end{adjustwidth}
\end{document}